\documentclass[10pt,journal]{IEEEtran}
\usepackage{amsmath,amsfonts}
\usepackage{amsthm}
\usepackage{algorithmic}
\usepackage{algorithm}
\usepackage{array}
\usepackage[caption=false,font=normalsize,labelfont=sf,textfont=sf]{subfig}
\usepackage{textcomp}
\usepackage{stfloats}
\usepackage{url}
\usepackage{verbatim}
\usepackage{graphicx}
\usepackage{cite}
\usepackage[pagewise]{lineno}
\usepackage{booktabs}
\usepackage{subfig}
\usepackage{balance}

\usepackage{multirow}
\usepackage{makecell}

\usepackage{xcolor}
\usepackage{xpatch}
\theoremstyle{definition}  
\newtheorem{remark}{Remark}

\begin{document}

\title{Joint Transmit and Pinching Beamforming Optimization in Pinching Antenna-Assisted Symbiotic Radio Systems}
\author{Ze Wang,~\IEEEmembership{Graduate Student Member,~IEEE}, Guoping Zhang,  Hongbo Xu, 

Ming Zeng, Fang Fang,~\IEEEmembership{Senior Member,~IEEE}
\thanks{
(Corresponding author: Hongbo Xu.)

Ze Wang, Guoping Zhang, Hongbo Xu are with the Department of Electronics and Information Engineering, Central China Normal University, Wuhan 430079, China (email: wangze0205@mails.ccnu.edu.cn; gpzhang@ccnu.edu.cn; xuhb@ccnu.edu.cn).

Ming Zeng is with the department of electrical and computer engineering, Laval University, Québec city, Canada, (email: ming.zeng@gel.ulaval.ca).

Fang Fang is with the Department of Electrical and Computer Engineering, Western University, London, ON N6A 5B9, Canada (e-mail: fang.fang@uwo.ca).

}}

\markboth{Journal of \LaTeX\ Class Files,~Vol.~18, No.~9, September~2024}%
{}

\maketitle

\begin{abstract}
This paper investigates a novel downlink symbiotic radio framework enabled by the pinching antenna system (PASS), designed to enhance both primary and secondary transmissions through reconfigurable antenna positioning. PASS consists of multiple waveguides equipped with low-cost pinching antennas, whose positions can be flexibly adjusted to jointly control large-scale path loss and signal phases. This reconfigurability introduces additional degrees of freedom for adaptive pinching beamforming, thereby enabling constructive signal enhancement and interference suppression tailored to the locations of the backscatter device, the internet-of-things (IoT) receiver, and the primary receivers. To fully exploit these benefits, we formulate a joint transmit and pinching beamforming optimization problem that maximizes the achievable sum rate while satisfying the IoT receiver’s detection error probability constraint and feasible deployment constraints for the pinching antennas. The resulting problem is inherently nonconvex and highly coupled. To address this challenge, we develop two complementary solution approaches. The first approach is a learning-aided gradient descent method, where the constrained optimization is reformulated into a differentiable form and solved through end-to-end learning. In this approach, the pinching antenna position matrix is reparameterized to automatically satisfy minimum spacing constraints, while transmit power and waveguide length limits are enforced via projection and normalization. The second approach is an optimization-based successive convex approximation–particle swarm optimization method, which first determines the transmit beamforming solution using successive convex approximation and subsequently optimizes pinching beamforming via a particle swarm optimization search over candidate pinching antenna placements. This two-stage approach achieves performance close to the element-wise optimal solution while significantly reducing computational complexity by restricting the search to an effective solution subspace. Furthermore, it demonstrates improved robustness compared to the learning-based method by mitigating the risk of convergence to undesirable local optima.
\end{abstract}


\section{Introduction}
\IEEEPARstart{W}{ith} the rapid advancement of the Internet of Things (IoT), the number of connected wireless devices is experiencing exponential growth and is projected to reach an estimated 5 trillion by 2030 \cite{9509294}. The realization of such a large-scale network necessitates the widespread deployment of IoT devices, which in turn imposes significant and unprecedented demands on both energy consumption and spectrum utilization. However, the limited available spectrum is far from sufficient to accommodate the rapidly growing demand from massive IoT deployments, particularly when each device requires a dedicated frequency band \cite{9040264}. Moreover, the widespread deployment of radio frequency (RF) components in IoT devices results in considerable energy consumption, thereby increasing the operational costs of large-scale networks. These challenges present major obstacles to the development of next-generation communication systems, highlighting the urgent need for innovative solutions that prioritize spectrum and energy efficiency \cite{9319204}. 

Recently, symbiotic radio (SR) has attracted increasing research interest due to its ability to enable spectrum- and energy-efficient IoT communications \cite{8907447}, offering a potential solution to the problems mentioned above. In SR, the passive secondary transmission, also referred to as IoT transmission \cite{9193946}, is achieved by the backscatter device (BD) acting, which operates as an IoT node by embedding its own information into the primary signal to IoT receiver (IR) without generating an RF carrier. Consequently, joint decoding can
be performed at the IR, allowing the backscatter transmission
to be decoded together with the primary signal, rather than
being treated as interference from the direct link as in AmBC
systems. Furthermore, the primary transmission can leverage
the backscatter transmission by treating the backscatter link
as an additional multipath component. Hence, they establish a mutually beneficial symbiotic relationship\cite{10045696}. Driven by these advantages, substantial research efforts have been devoted to SR\cite{8907447,9652042,8638762}. Based on the relationship between the symbol durations of the primary transmitter (PT) and the BD, SR can be categorized into commensal SR (CSR) and parasitic SR (PSR). In PSR, the BD and PT share the same symbol duration, whereas in CSR, the BD's symbol duration is significantly longer than that of the PT \cite{8907447,9652042}. To enhance spectral efficiency, transmit power minimization has been investigated for full-duplex SR systems \cite{8638762}. Although BD can provide an additional reflection path, the assistance from the reflection path alone is very weak, when the line-of-sight (LoS) path experiences severe fading or blockage. Improving wireless environments has become a key research focus, aiming to create favorable channel conditions to enhance communication performance\cite{10923651}. 

\subsection{Related Work}
\subsubsection{Reconfigurable Antenna aided SR}
To enable efﬁcient backscatter communication, \cite{9769767} investigated the roles of reconﬁgurable intelligent surfaces (RIS) for assisting different scenarios of backscatter communication. By configuring RIS to act as a BD, backscatter communication can benefit from the additional spatial degrees of freedom introduced by multiple reflecting elements. Furthermore, the total transmit power minimization problem in RIS-assisted multiple-input multiple-output (MIMO) SR systems has been analyzed in \cite{9391685}. In \cite{11060576}, RIS-enabled SR with orthogonal frequency division multiplexing (OFDM) transmission has been investigated under imperfect symbol synchronization to improve spectral efficiency. More recently, movable antennas (MAs) \cite{10286328}, also known as fluid antennas \cite{9264694}, have emerged as a promising technology for intelligently reconfiguring wireless environments by flexibly adjusting antenna positions. The utilization of MAs in PT can effectively improve the rate of secondary transmission by optimizing the positions of MAs to strengthen the beamforming gain at the BD \cite{10636790,10683667}. However, the performance improvements offered by these technologies are often limited due to severe path loss, particularly in high-frequency bands \cite{9267779}. For instance, the double fading effect inherent in the cascade reflection link of RIS leads to significantly higher path loss compared to a direct LoS link \cite{8936989}. Similarly, the movement range of MAs is typically limited to only a few wavelengths, which restricts their overall performance gains.

\subsubsection{Pinching-Antenna Systems}

To overcome these limitations, the pinching antenna system (PASS), recently proposed by DOCOMO \cite{suzuki2022pinching}, has been recognized as a potential solution in the domain of flexible-antenna technologies \cite{suzuki2022pinching,10945421,zeng2025resource}. PASS leverages a dielectric waveguide as its transmission medium to establish adjustable LoS links with users. The system enables the signal radiation from any desired radiation points that are activated by implementing dielectric particles \cite{wang2025model}. These dielectric particles are referred to as pinching antennas (PAs), which exhibit properties similar to those of leaky-wave antennas \cite{karagiannidis2025}. However, in contrast to the leaky-wave based systems where the antennas are fixed in place with pre-defined locations, PAs support flexible and dynamic activation, allowing signals radiated from dielectric waveguides to adapt effectively to complex and time-varying environments. This capability enables a cost-efficient and scalable MIMO implementation through the novel concept of pinching beamforming \cite{liu2025pinching}, which enhances communication performance by dynamically optimizing antenna configurations.

Driven by the above promising characteristics, PASS has attracted increasing research attention, although it remains in the early stages of development. In \cite{10945421}, both single-waveguide and multi-waveguide scenarios were investigated, and low-complexity pinching beamforming schemes were proposed for single-user and two-user multi-input single-output (MISO) systems. In \cite{xu2025join}, the authors addressed a joint transmit and pinching beamforming optimization problem for a multi-user PASS downlink framework, introducing both an optimization-based majorization-minimization and penalty dual decomposition method and a learning-based knowledge-guided dual learning approach. Furthermore, the authors in \cite{11018390} developed two efficient deep learning-driven channel estimation methods for PASS, demonstrating their superior estimated performance and low pilot overhead. The achievable array gain of PASS was analyzed in \cite{10981775, 11036558 }, where \cite{10981775} proposed an antenna position enhancement algorithm to approximate its performance upper bound, and \cite{11036558 } demonstrated that LoS blockage can enhance the performance advantage of pinching antennas over conventional antennas. Additionally, energy-efficient resource allocations for PASS were studied in \cite{11123791} and \cite{11131179}. Moreover, a comprehensive analytical framework is introduced for evaluating PASS performance in \cite{10976621}, with closed-form expressions derived for the average achievable rate and outage probability.

\subsection{Motivations and Contributions}

Based on the above discussion, PASS has demonstrated strong capabilities in establishing robust LoS links, significantly reducing free-space propagation loss, and overcoming blockage issues. Consequently, the utilization of PAs in the SR systems is essential for achieving highly reliable and spectrally efficient primary and secondary transmissions. Moreover, different from conventional antenna systems, the flexible deployment of PAs introduces additional degrees of freedom, facilitating effective pinching beamforming tailored to the locations of the BD, the IR, and primary receivers (PR). To the best of our knowledge, the application of PASS in SR systems remains largely unexplored in the existing literature. Motivated by this gap, this paper proposes a PASS-enabled downlink SR framework and develops joint beamforming methods. The main contributions of this work are summarized as follows.

\begin{itemize}
\item[$\bullet$] We propose a novel PASS-assisted downlink SR framework, where the PASS BS with multiple waveguides treated as a PT serves an IR and PRs with the assistance of the BD. Within this model, we formulate a joint transmit and pinching beamforming optimization problem for maximizing the sum rate, while satisfying the constraints of the detection error probability of the IR and the feasible deployment region of PAs. To tackle this highly coupled nonconvex problem, we develop both learning-aided gradient descent (LGD) and two-stage optimization-based algorithms.

\item[$\bullet$] For the LGD algorithm, we address the constraints by equivalently transforming the constrained optimization problem into a tractable form that can be directly solved using gradient descent. Specifically, we reparameterize the position matrix of PAs as non-negative offsets to satisfy the minimum spacing constraint, while the maximum transmit power and waveguide length constraints are handled via projection and normalization techniques. Conventional manual gradient derivation or symbolic differentiation often leads to expression swelling and computational inefficiency. In contrast, the proposed LGD framework leverages automatic differentiation and the Adam optimizer, allowing efficient updates of optimization variables modeled as learnable parameters updated by back-propagation.

\item[$\bullet$] For the two-stage optimization-based approach, we utilize a successive convex approximation and particle swarm optimization (SCA-PSO) algorithm. The original joint optimization problem is decoupled into two subproblems. In the first stage, we approximate the subproblem with respect to the transmit beamforming matrix, which is inherently non-convex, via SCA by transforming the objective and constraint functions into concave forms, enabling efficient solution through convex optimization tools. In the second stage, we address the pinching beamforming design using a PSO-based algorithm, where each particle encodes a candidate PA deployment matrix, and its fitness is evaluated based on the resulting achievable sum rate.

\item[$\bullet$] Finally, numerical simulations are conducted to evaluate the effectiveness of the proposed framework and algorithms. The results demonstrate that: i) The proposed PASS-enabled SR system achieves significantly higher sum rates compared to conventional antenna-based SR schemes. ii) The proposed SCA-PSO algorithm improves performance by 17.1\% and 35.5\% over the low-complexity LGD and fixed-PA schemes, respectively, and approaches the performance of the element-wise optimization method.
\end{itemize}

\subsection{Organization and Notations}

The structure of the paper is as follows. Section II describes the PASS-assisted SR system and formulates the sum-rate maximization problem. Section III introduces the equivalent reformulation of the original problem and proposes a GD-based joint beamforming framework. Section IV presents the proposed SCA-PSO algorithm for joint transmit and pinching beamforming, designed to further enhance system performance. Section V provides numerical results that validate the convergence behavior and demonstrate the performance advantages of the proposed framework and algorithms. The concluding remarks are provided in Section VI.

Notation: Scalars, vectors, and matrices are represented by $x$, $\mathbf{x}$, and $\bold X$, respectively. ${\left( \cdot \right)^T}$, ${\left( \cdot \right)^*}$, and ${\left( \cdot \right)^H}$ stand for the transpose, complex conjugate, and conjugate transpose operations, respectively. The notation ${\mathop{\rm Re}\nolimits} \left\{ \cdot \right\}$ and ${\mathop{\rm Im}\nolimits} \left\{ \cdot \right\}$ denote the real and imaginary part of a complex number, respectively. ${\rm{Tr}}\left( \cdot \right)$, $\left| \cdot \right|$, and $\left\| \cdot \right\|$ represent the trace, the modulus operator, and the Euclidean norm, respectively. The blkdiag $(a_1,...,a_N)$ is a block diagonal matrix with diagonal blocks $a_1,...,a_N$. ${C^{M \times N}}$ denotes the dimension of an $M \times N$ complex-valued matrix. ${\cal C}{\cal N}\left( {\mu ,\sigma^2} \right)$ is the circularly symmetric complex Gaussian random distribution with mean $\mu $ and variance $ \sigma^2 $.

\section{System Model}
\begin{figure}[t]
\centering
\includegraphics[width=0.49\textwidth]{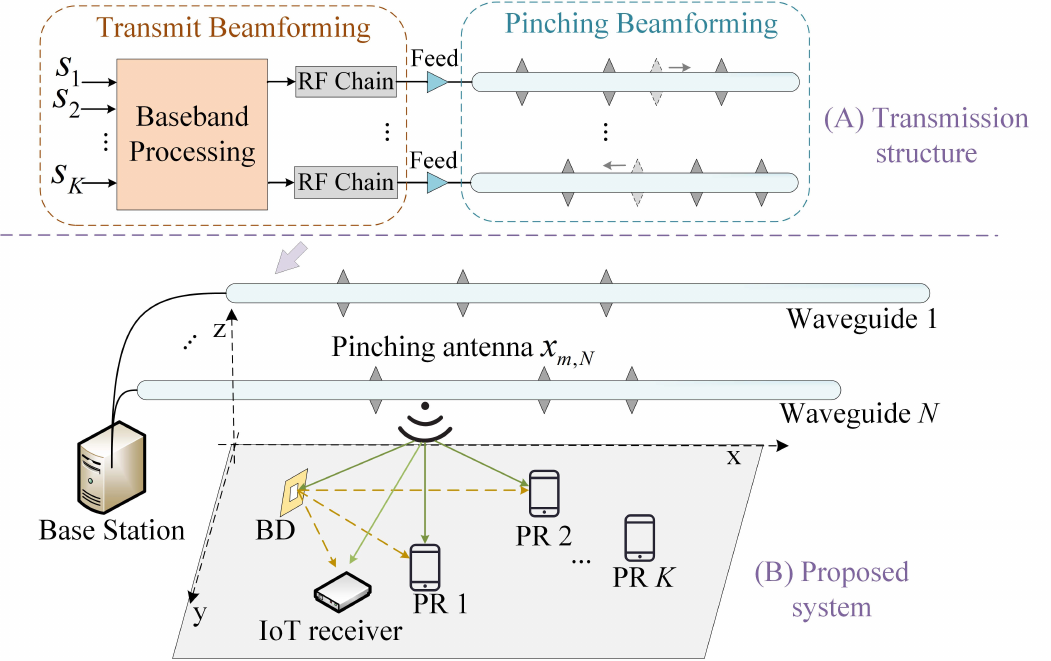}
\caption{Illustration of the considered downlink PASS-assisted SR system. (A) presents the joint transmit and pinching beamforming architecture. (B) describes that the BS employs $N$ waveguides, each integrated with $M$ PAs, to simultaneously serve $K$ PRs and an IR via the BD.}
\label{Fig.1}
\end{figure}

As illustrated in Fig. \ref{Fig.1}, this paper considers a downlink PASS-assisted SR system, where the PASS is connected to a base station (BS) to simultaneously serve $K$ single-antenna PRs and a single-antenna IR via a BD. The PASS consists of $N$ dielectric waveguides, with $N \ge  K$, where each waveguide is incorporated with $M$ pinching antennas \cite{xu2025join}. Activating PAs at different positions along the waveguides enables flexible control over the phases of the incident signals and the large-scale fading. The waveguides in PASS are fed by the BS, which transmits primary signals to PRs with BD-assisted backscattering. This process not only facilitates primary transmission but also embeds its messages within the primary signals intended for the IR.

\subsection{Channel Model}

Assume that both PAs and waveguides are located at a fixed height of $z^{\rm{PA}}$, and a three-dimensional Cartesian coordinate system is established. The location of the $m$-th PA associated with the $n$-th waveguide is denoted by ${\mathbf I}_{n,m}^{{\rm{PA}}} = {[{x_{n,m}},{y_n},{z^{{\rm{PA}}}}]^T}$, where $ {x_{n,m}}$ is the adjustable coordinate over the $x$-axis, and $y_n$ the fixed and pre-defined coordinate over $y$-axis, and ${\mathbf I}_k^{\rm{U}} = {[x_k^{\rm{U}},y_k^{\rm{U}},0]^T}$ represent the $k$-th PR’s position. Given that non-line-of-sight (NLoS) paths are significantly weaker than line-of-sight (LoS) paths, we adopt a practical channel model that considers only the LoS components while ignoring the NLoS components \cite{8901159,11090043}. Based on the geometric free-space spherical model, the channel from the PA $ {\mathbf{ I} _{n,m}^{{\rm{PA}}}}$ to the $k$-th PR at the $ \mathbf I _k^{\rm{U}} $ is given by

\begin{equation}
h_{k,n,m}^H({x_{n,m}}) = \frac{{\kappa {e^{ - j{\beta _h}r({x_{n,m}},{\mathbf I}_k^{\rm{U}})}}}}{{r({x_{n,m}},{\mathbf I}_k^{\rm{U}})}}, \label{eq1}
\end{equation}
where $\kappa  = c/4\pi {f_c}$ denotes the reference channel gain at a distance of 1 m, with $c$ and $f_c $ representing the speed of light and carrier frequency, respectively.  ${\beta _h} = {{2\pi } \mathord{\left/
 {\vphantom {{2\pi } {{\lambda _f}}}} \right.
 \kern-\nulldelimiterspace} {{\lambda _f}}}$ is the wave number in the propagation medium, and ${\lambda _f}$ is the corresponding wavelength.  $r({x_{n,m}},{\mathbf I}_k^{\rm{U}}) = \left\| {{\mathbf I}_k^{\rm{U}} - {\mathbf I}_{n,m}^{{\rm{PA}}}} \right\|$ represents the distance from the PA $ {\mathbf I _{n,m}^{{\rm{PA}}}}$ to $ {{\mathbf I}_k^{\rm{U}}}$. Similarly, the channels from the PA $ {\mathbf I _{n,m}^{{\rm{PA}}}}$ to the IR and the BD are denoted by $h_{{\text{\text{IR}}},n,m}^H$ and $h_{\text{\text{BD}},n,m}^H$, respectively. Stacking the channel vectors from all the PAs to the $k$-th PR, $\mathbf{h}_k^H(\bold X) = [\mathbf{h}_{k,1}^H({x_1}),\ldots,\mathbf{h}_{k,N}^H({x_N})] \in {\mathbb{C}^{1 \times NM}}$ is the overall channel vector for the $k$-th PR. Furthermore, for in-waveguide transmission, we denote the diagonal matrix $\bold{G}(\bold{X}) \in {\mathbb{C}^{NM \times N}}$ as representing the path response from the feed point of each waveguide to the corresponding PAs, which is given by
\begin{align}
\nonumber
{\bf{G}}({\bf{X}})& = {\text{blkdiag}}({\bf{g}}\left( {{x_1}} \right),...,{\bf{g}}\left( {{x_N}} \right)) \\
\nonumber
&= \left[ {\begin{array}{*{20}{c}}
{{\bf{g}}\left( {{x_1}} \right)}&0& \cdots &0\\
0&{{\bf{g}}\left( {{x_2}} \right)}& \cdots &0\\
 \vdots & \vdots & \ddots & \vdots \\
0&0& \cdots &{{\bf{g}}\left( {{x_N}} \right)}
\end{array}} \right].
\end{align}
The response vector $ \mathbf{g}\left( {{x_1}} \right)$ is characterized by
\begin{equation}
\mathbf{g}\left( {{x_n}} \right) = [{\upsilon _1}{e^{ - j{\beta _g}{x_{n,m}}}},\ldots,{\upsilon _M}{e^{ - j{\beta _g}{x_{n,M}}}}], \label{eq2}
\end{equation}
where ${\upsilon _1}$ is the amplitude of the transmitted signal and ${\beta _g} = {{2\pi {n_{\text{eff}}}} \mathord{\left/
 {\vphantom {{2\pi n_{\text{eff}}} {{\lambda _f}}}} \right.
 \kern-\nulldelimiterspace} {{\lambda _f}}}$ denotes the propagation constant of the waveguide, with ${{n_{\text{eff}}}}$ being the effective refractive index of the dielectric waveguide.

\subsection{Signal Model}
Denote the symbol vector transmitted from the BS to the PRs  by $\mathbf{s}(l) \in {\mathbb{C}^{K \times 1}}$ with $\mathbb{E}[\mathbf{s}(l){\mathbf{s}^H}(l)] = {\bold{I}_K}$. Let $\bold{W} = [{\mathbf{w}_1},\ldots,{\mathbf{w}_K}] \in {\mathbb{C}^{N \times K}}$ represent the transmit beamforming matrix, with the total transmit power satisfying $\operatorname{Tr}({\bold{W}}{{\bold{W}}^H}) \le {P_{\max }}$. Subsequently, the transmitted signal at the BS is given by ${\bold{W}}\mathbf{s}\left( l \right)$. Additionally, the BD transmits its own signal $c$ to the IR by using the ON-OFF keying (OOK) modulation, i.e., $c = $ “0” and “1” correspond to OFF and ON states, respectively. At the $l$-th time slot, the received signal at the $k$-th PR is given by\protect\footnotemark,
\footnotetext{In this work, we assume perfect channel state information (CSI) to analyze the fundamental performance limits, while practical channel estimation methods can be found in \cite{liu2025pinching,11018390}. Efficiently estimating CSI in PASS remains a topic for future research. }

\begin{equation}
{y_k}(l) = \left( {\mathbf{h}_k^{eq} + c{f_{\text{BD},k}}\mathbf{h}_{\text{BD}}^{eq}} \right){\bold W}\mathbf{s}(l) + {n_k}(l), \label{eq3}
\end{equation}
where $\mathbf{h}_k^{eq} = \mathbf{h}_k^H(\bold X)\bold{G}(\bold{X})$ and $\mathbf{h}_{\text{BD}}^{eq} = \mathbf{h}_{\text{BD}}^H(\bold{X})\bold{G}(\bold{X})$ are the equivalent channels from the BS to the $k$-th PR and BD, respectively. ${{f_{\text{BD},k}}}$ and ${n_k}(l) \sim {\cal C}{\cal N}(0,\delta _k^2)$ denote the channel reflective-link from the BD to the $k$-th PR and the additive white Gaussian noise, while $ {{{P_{\max }}} \mathord{\left/
 {\vphantom {{{P_{\max }}} {\delta _k^2}}} \right.
 \kern-\nulldelimiterspace} {\delta _k^2}}$ is the transmit signal-to-noise ratio (SNR) of the $k$-th user \cite{10945421}. Since the communication rate of $c$ is much lower than that of $\mathbf{s}(l)$, we assume $T_c = L T_s$, $L \gg 1$, where $T_c$ and $T_s$ denote the symbol period of $c$ and $\mathbf{s}(l)$. When decoding $\mathbf{s}(l)$, the backscatter link formed by the BD can be treated as an additional path. Given that the PRs possess no prior information regarding the BD’s symbol $c$, we assume non-coherent detection can be applied to detect $\mathbf{s}(l)$ with partial CSI \cite{1362900}. Therefore, the signal-to-interference-plus-noise ratio (SINR) at the $k$-th PR can be expressed as

\begin{equation}
{\text{SINR}_k} = {\frac{{{{\left| {\left( {\mathbf{h}_k^{eq} + c{\mathbf{f}_{b,k}}} \right){\mathbf{w}_k}} \right|}^2}}}{{\sum {_{i = 1,i \ne k}^K{{\left| {\left( {\mathbf{h}_k^{eq} + c{\mathbf{f}_{b,k}}} \right){\mathbf{w}_i}} \right|}^2} + \delta _k^2} }}}, \label{eq4}
\end{equation}
where ${\mathbf{f}_{b,k}} = {f_{\text{BD},k}}\mathbf{h}_{\text{BD}}^{eq}$ denote the backscatter cascade channel from the BS to the $k$-th PR via the BD. Assuming that the BD transmits the symbols “0” and “1” with equal a priori probability, the average achievable rate of decoding $\mathbf{s}(l)$ at the $k$-th PR is expressed as \cite{9391685}

\begin{align}
\nonumber
{R_k}& = {\mathbb{E}_c}\left[\log(1 + {\text{SINR}_k}) \right] \\
\nonumber
&= \frac{1}{2}{\log _2}(1 + \frac{{{{\left| {\mathbf{h}_k^{eq}{\mathbf{w}_k}} \right|}^2}}}{{\sum {_{i = 1,i \ne k}^K{{\left| {\mathbf{h}_k^{eq}{\mathbf{w}_i}} \right|}^2} + \delta _k^2} }}) \\
&\quad + \frac{1}{2}{\log _2}(\frac{{{{\left| {(\mathbf{h}_k^{eq} + {\mathbf{f}_{b,k}}){\mathbf{w}_k}} \right|}^2}}}{{\sum {_{i = 1,i \ne k}^K{{\left| {(\mathbf{h}_k^{eq} + {\mathbf{f}_{b,k}}){\mathbf{w}_i}} \right|}^2} + \delta _k^2} }}). 
\end{align}

At the $l$-th time slot within one BD symbol period, the received signal at the IR is given by

\begin{equation}\label{eq6}
{y_{\text{IR}}}(l) = \left( {h_{\text{IR}}^{eq} + c{f_{\text{b},\text{IR}}}} \right){\bold{W}}\mathbf{s}(l) + {n_{\text{IR}}}(l),
\end{equation}
where ${\mathbf{f}_{\text{b},\text{IR}}} = {f_{\text{BD},{\text{IR}}}}\mathbf{h}_{\text{BD}}^{eq}$. The IR aims to recover the symbol $c$ from the received signal by distinguishing between two hypotheses corresponding to the BD's transmitted symbol, either “0” or “1”. In the received signal at the IR, the direct link component is much stronger than the backscatter due to the double fading effect over the backscattering link, and a successive interference cancellation (SIC) scheme is designed to recover the data symbol of the BD \cite{8907447,9481926}. Specifically, the IR first decodes the primary signal $\mathbf{s}(l)$ by applying maximum ratio combining (MRC). After successfully decoding $\mathbf{s}(l)$, the IR subtracts direct-link interference$\mathbf{h}_{\text{IR}}^{eq}{\bold W}\mathbf{s}(l)$ from the received signal (6). Subsequently, the two hypotheses are represented by
\begin{equation}
{\bar y_{\text{IR}}}(l) \leftarrow \left\{ {\begin{array}{*{20}{c}}
{{{n_{\text{IR}}}(l)},\qquad \qquad {\quad}{{H_0}}}\\
{{{\mathbf{f}_{\text{b},\text{IR}}} {\bold{W}}\mathbf{s}(l) + {n_{\text{IR}}}(l)},{} H_1}
\end{array}}, \right. \label{eq7}
\end{equation}where the OFF and ON states in OOK are associated with the null hypothesis $H_0$ and alternative hypothesis $H_1$, respectively. The detection performance at the IR is then evaluated in terms of the detection error probability, which is expressed as

\begin{equation}\label{eq8}
\xi  = \Pr \left( {{{\cal B}_1}\left| {{H_0}} \right.} \right) + \Pr \left( {{{\cal B}_0}\left| {{H_1}} \right.} \right)
\end{equation}
where  $\Pr \left( {{{\cal B}_1}\left| {{H_0}} \right.} \right)$ and $\Pr \left( {{{\cal B}_0}\left| {{H_1}} \right.} \right)$ denote the false alarm rate and miss detection rate, respectively. ${{\cal B}_1}$ and ${{\cal B}_0}$ represent the binary decisions that determine whether the backscatter link is present or not, respectively. Based on the Neyman-Pearson criterion, the likelihood ratio test is employed to minimize the detection error probability $\xi$ \cite{9496108}, and is formulated as

\begin{equation}\label{eq9}
\frac{{{P_1} \buildrel \Delta \over = \prod {_{l = 1}^Lf({{\bar y}_{\text{IR}}}(l)\left| {{H_1}} \right.)} }}{{{P_0} \buildrel \Delta \over = \prod {_{l = 1}^Lf({{\bar y}_{\text{IR}}}(l)\left| {{H_0}} \right.)} }}{\rm{ }}\frac{{\mathop  > \limits^{{{\cal B}_0}} }}{{\mathop  < \limits_{{{\cal B}_1}} }}1.
\end{equation}
The likelihood functions of ${\bar y_{\text{IR}}}(l)$ in $H_0$ and $H_1$ are denoted as $f({\bar y_{\text{IR}}}(l)\left| {{H_0}} \right.) \sim CN(0,\delta _{{\rm{\text{IR}}}}^2)$ and $f({\bar y_{\text{IR}}}(l)\left| {{H_1}} \right.) \sim CN(0,{\gamma _b} + \delta _{{\rm{\text{IR}}}}^2)$ with ${\gamma _b} = {\left\| {{f_{\text{BD},{\text{IR}}}}\mathbf{h}_{\text{BD}}^{eq}{\bold{W}}} \right\|^2}$, respectively. Then, the minimum detection error rate $P_e$ can be derived from (\ref{eq8}) and (\ref{eq9}). However, as the resultant expression of $P_e$ involves the incomplete Gamma function, it poses challenges to further analytical and design efforts\cite{9496108}. To address this, a tractable lower bound on $P_e$ is obtained according to \cite{6584948}, expressed as follows

\begin{equation}\label{eq10}
{P_e} \le 1 - \sqrt {\frac{1}{2}{\cal D}\left( {{P_0}\left\| {{P_1}} \right.} \right)},
\end{equation}
where ${\cal D}\left( {{P_0}\left\| {{P_1}} \right.} \right) = L\left[ {\ln (\frac{{{\gamma _b} + \delta _{{\rm{\text{IR}}}}^2}}{{\delta _{{\rm{\text{IR}}}}^2}}) + \frac{{\delta _{{\rm{\text{IR}}}}^2}}{{{\gamma _b} + \delta _{{\rm{\text{IR}}}}^2}} - 1} \right]$ denotes Kullback-Leibler (KL) divergence from $P_0$ to $P_1$. Hence, the detection constraint for the secondary transmission is derived as ${\cal D}\left( {{P_0}\left\| {{P_1}} \right.} \right) \ge 2{\varepsilon ^2}$, which is a more stringent constraint to guarantee ${P_e} \le 1 - \varepsilon $.

\subsection{Problem Formulation}
This paper jointly optimizes the transmit beamforming at the BS and the pinching beamforming formed by the PAs, with the aim of maximizing the sum rate of PRs. We formulate the corresponding optimization problem as

\begin{subequations}\label{eq11}
\begin{align*}
(\text{P1})\mathop {\max }\limits_{\bold{W}, \bold{X}}\, &\sum\limits_{k = 1}^K {{R_k}} \tag{11a}\label{11a}\\
{\rm{s.t.}} &\operatorname{Tr}({\bold{W}}{{\bold{W}}^H}) \le {P_{\max }}, \tag{11b}\label{11b}\\
&D({P_0}|{P_1}) \ge 2{\varepsilon ^2}, \tag{11c}\label{11c}\\
&{x_{n,m + 1}} - {x_{n,m}} \ge {d_{\min }},\forall n,m, \tag{11d}\label{11d}\\
&0 \le {x_{n,m}} \le {S_x},\forall n,m, \tag{11e}\label{11e}
\end{align*}
\end{subequations}
where (\ref{11b}) represents the maximum transmit power constraint, (\ref{11c}) ensures the minimum detection error rate at the IR, (\ref{11d}) imposes a minimum antenna spacing $d_{\text{min}}$ to avoid mutual coupling between adjacent PAs, and (\ref{11e})  guarantees that the positions of the PAs are within the maximum range of the connected waveguide. 

The optimization problem (P1) is highly non-convex and intractable due to the fractional expressions and multivariable coupling in both the objective function and the constraints. In the following section, we first introduce a learning-based beamforming framework to solve problem (P1). Furthermore, an alternating optimization method is employed to decompose the original problem into two sub-problems, which are then solved iteratively using SCA and PSO methods.

\section{Gradient Descent-based Joint Beamforming Framework}
In this section, we propose an LGD method to solve the joint beamforming problem (P1). Specifically, we first address the constraints and equivalently transform the constrained optimization problem into an unconstrained form. Subsequently, we present the proposed LGD-based beamforming design algorithm, which is implemented using self-defined neural network layers, where the optimization variables are treated as learnable parameters.

\subsection{GD-based Reformulation}
\begin{figure}[htbp]
\centering
\includegraphics[width=0.48\textwidth]{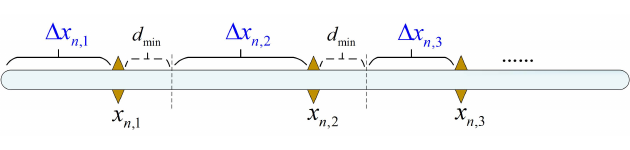}
\caption{Illustration of mapping from the PA positions to the offsets $ \Delta {x_{n,m}}$, $m=1,…,M$.}
\label{Fig.2}
\end{figure}

A major challenge in designing the pinching beamforming lies in efficiently handling constraints (\ref{11d}) and (\ref{11e}). To overcome this, we reformulate the original optimization variable $\mathbf{x}_n$ into a more tractable form by introducing an offset variable $\Delta \mathbf{x}_n$. Specifically, inspired by the minimum distance constraints of PAs, constraint (\ref{11d}) can be reformulated as 
${x_{n,m + 1}} \ge {x_{n,m}} + {d_{\min }} \to {x_{n,m + 1}} = {x_{n,m}} + {d_{\min }} + \Delta {x_{n,m + 1}}$, where $ \Delta {x_{n,m + 1}}$ denotes the non-negative offset of the $(m+1)$-th PA relative to $m$-th PA at the $n$-th waveguide. Hence, we define the position of the first PA ${x_{n,1}} = \Delta {x_{n,1}},\Delta {x_{n,1}} \ge 0$, and constraint (\ref{11d}) can be further rewritten as

\begin{equation}\label{eq12}
\left\{ 
\begin{aligned}
&x_{n,1} = \Delta x_{n,1}, \\
&x_{n,m} = (m - 1){d_{\min }} + \sum_{j = 1}^m \Delta x_{n,j},\quad m = 2,\ldots,M
\end{aligned}
\right.
\end{equation}
for $\forall n$. For clarity, Fig. \ref{Fig.2} illustrates the parameter mapping of the PA position matrix in the $n$-th waveguide, with the offsets $ \Delta {x_{n,m}}$, $m=1,…,M$, being defined as the optimization variables.

However, $\left\{ {\Delta {x_{n,m}}} \right\}_{m = 1}^M$ are also restricted by the maximum length constraint (\ref{11e}), such as ${x_{n,M}} \le {S_x}$. which can be equivalently written as

\begin{equation}\label{eq13}
\sum {_{j = 1}^m\Delta {x_{n,j}}} \le {\Delta _{\max }},
\end{equation}
where ${\Delta _{\max }} = {S_x}  - (m - 1){d_{\min }}$. When this constraint is not satisfied, the optimized $\Delta {\mathbf{x}_n}$ will be normalized through the softmax function, i.e., $f_{\text{SM}}(\Delta {x_{n,m}}) = \frac{{\Delta {x_{n,m}}}}{{\sum {_{i = 1}^M\Delta {x_{n,i}}} }}{\Delta _{\max }}$. This normalization ensures that the resulting offsets satisfy the maximum aperture constraint.  Accordingly, the position optimization of $x_n$ can be implemented by optimizing the corresponding offset vector $\Delta \mathbf{x}_n$, which efficiently handles the constraints (\ref{11d}) and (\ref{11e}).

For the transmit beamforming, the projected gradient descent method \cite{9837315} is adapted to deal with the maximum transmit power constraint (\ref{11b}). Let the constraint set be denoted by ${\cal C} \buildrel \Delta \over = \left\{ {\left. {\bold{W}} \right|\operatorname{Tr}({\bold{W}}{{\bold{W}}^H}) \le {P_{\max }}} \right\}$, the projection operation $\prod {_{\cal C}} $ can be expressed as

\begin{equation}\label{eq14}
\prod {_{\mathcal C}} \{ W\} = 
\begin{cases}
{\bold{W}}, & \text{if } \operatorname{Tr}({\bold{W}} {\bold{W}}^H) \le P_{\max}. \\
\frac{{\bold{W}}}{\| {\bold{W}} \|}\sqrt{P_{\max}}, & \text{otherwise}.
\end{cases}
\end{equation}
Moreover, following \cite{10741192}, the penalty method can be adopted to guarantee the constraint (\ref{11c}). Specifically, we introduce a penalty parameter $\xi$ to the objective function (\ref{11a}), allowing the original problem to be reformulated as follows:

\begin{equation}\label{eq15}
(\text{P1.1})\mathop {\min }\limits_{\bold{W}, \bold{X}} {\cal F} \buildrel \Delta \over = -\sum\limits_{k = 1}^K {{R_k}}  + \xi {[\max (0,2{\varepsilon ^2} - D({P_0}\left| {{P_1}} \right.))]^2}
\end{equation}
where $\xi$ controls the penalty magnitude. It is important to note that the objective function $ {\cal F}$ is differentiable, and its gradient with respect to the transmit beamforming ${\bold{W}}$ can be denoted as

\begin{equation}\label{eq16}
{\nabla _{\bold{W}}}{\cal F} = \sum\limits_{k = 1}^K {{\nabla _{\bold{W}}}{R_k}}  - 2\xi \frac{{\partial D({P_0}|{P_1})}}{{\partial {\gamma _b}}} \cdot \frac{{\partial {\gamma _b}}}{{\partial {\bold{W}}}}.
\end{equation}
Thus, the update of ${\bold{W}}$ is obtained by

\begin{equation}\label{eq17}
{\widetilde {\bold{W}}^{(i)}} = {\bold{W}}^{(i-1)} - {\eta _1}{\nabla _{\bold{W}}}{\cal F}
\end{equation}
\begin{equation}\label{eq18}
{{\bold{W}}^{(i)}} = \prod {_{\cal C}} ({\widetilde {\bold{W}}^{(i)}})
\end{equation}
where ${\eta _1}$ is the step size. Then, the gradient vector for the objective function with respect to the pinching beamforming $ \Delta {\mathbf{x}_n}$ is given by

\begin{equation}\label{eq19}
{\nabla _{\Delta {\mathbf{x}_n}}}{\cal F} = {[\frac{{\partial {\cal F}}}{{\partial \Delta {x_{n,1}}}},\frac{{\partial {\cal F}}}{{\partial \Delta {x_{n,2}}}},\ldots,\frac{{\partial {\cal F}}}{{\partial \Delta {x_{n,M}}}}]^T},\forall n.
\end{equation}
Notably, the derivation of ${\cal F}$ involves the chain rule, rendering a closed-form analytical expression intractable. Instead, by computing the gradient $ {\nabla _{\Delta {\mathbf{x}_n}}}{\cal F}$, the variable ${\Delta {\mathbf{x}_n}}$ can be updated by

\begin{equation}\label{eq18}
\widetilde \Delta {\mathbf{x}_n}^{(i)} = \Delta {\mathbf{x}_n}^{(i - 1)} - {\eta _2}{\nabla _{\Delta {\mathbf{x}_n}}}{\cal F},
\end{equation}
\begin{equation}\label{eq19}
\Delta {\mathbf{x}_n}^{(i)} = {f_{\text{SM}}}\left( {\widetilde \Delta {\mathbf{x}_n}^{(i)}} \right),
\end{equation}
where ${\eta _2}$ also denote the step size. After obtaining the optimized $\Delta {\mathbf{x}_n}$, we can recover ${{x_{n}}}$ from (\ref{eq12}).

\begin{figure*}[t]
\centering
\includegraphics[width=0.65\textwidth]{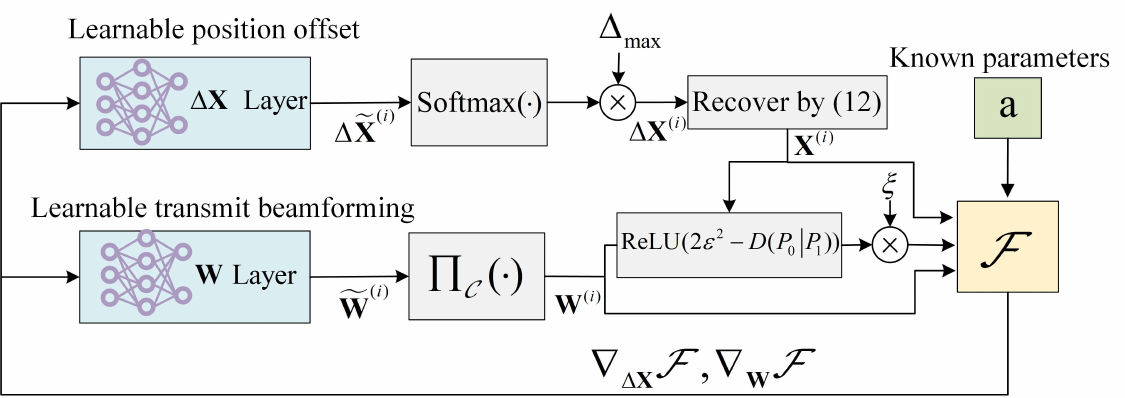}
\caption{Learning-aided GD for joint beamforming in the PASS-enabled SR system. The learnable parameter matrices of the $\Delta \bold{X} $ layer (position offsets of the PAs) and the ${\bold{W}}$ layer (transmit beamforming matrix) are jointly optimized through iterative back-propagation of the loss function $ {\cal F}$. Gray-shaded blocks indicate the operations of handling constraints, and $\mathbf{a} $ denotes the known channel spatial location information.}
\label{Fig.3}
\end{figure*}

\subsection{Learning-aided GD Approach}
In conventional GD, computing and projecting the gradients as described in (16) and (19) can be challenging and often intractable. Moreover, selecting appropriate fixed step sizes $\left\{ {{\eta _1},{\eta _2}} \right\}$ is nontrivial and may significantly affect convergence performance. 

In this paper, we propose a learning-aided gradient descent (LGD) framework that follows the fundamental principle of the classical GD algorithm. Specifically, the proposed LGD introduces two self-defined layers in which the optimization variables are treated as learnable parameters and iteratively updated through back-propagation. In practice, the parameter updates are implemented using the Adam optimizer, which provides adaptive step sizes and momentum to improve convergence stability. Furthermore, the proposed LGD framework leverages automatic differentiation, enabling efficient and broadly applicable gradient descent solutions for practical optimization tasks \cite{9805773}.

Fig. \ref{Fig.3} illustrates the architecture of the proposed LGD for joint beamforming, where $\Delta \bold{X} $ layer and ${\bold{W}}$ layer are implemented\protect\footnotemark,
\footnotetext{The transmit beamforming matrix $\bold W$ and the equivalent channel vector $ \mathbf{h}_k^{eq}(\Delta \bold{X})$ are complex-valued. However, optimizing complex-valued parameters directly in neural networks is not straightforward. To address this issue, we introduce two self-defined network layers \cite{9707491} with dimension $ N \times K$, which correspond to ${\rm{Re}}\left\{ \bold W \right\}$ and ${\rm{Im}}\left\{ \bold W \right\}$. The complex matrix-vector product $\mathbf{h}_k^{eq}\bold{W}$ is then reformulated as an equivalent real-valued linear operation, as given in
$\left[ {\begin{array}{*{20}{c}}
{{\mathop{\rm Re}\nolimits} (\bold W)}&{ - {\mathop{\rm Im}\nolimits} (\bold W)}\\
{{\mathop{\rm Im}\nolimits} (\bold W)}&{{\mathop{\rm Re}\nolimits} (\bold W)}
\end{array}} \right]\left[ {\begin{array}{*{20}{c}}
{{\mathop{\rm Re}\nolimits} (\mathbf{h}_k^{eq})}\\
{{\mathop{\rm Im}\nolimits} (\mathbf{h}_k^{eq})}
\end{array}} \right]$, where $\Delta \bold{X} $ layer and ${\bold{W}}$ layer contain $NM$ and $2NK$ learnable parameters, respectively.}with their parameters corresponding to the position offsets of PAs and the transmit beamforming matrix, both of which are iteratively updated at each step via back-propagation of the loss function $ {\cal F}$. 
Note that the channel location information of the BD, IR, and PRs constitutes the known parameter set $\mathbf{a} = \{ \mathbf{I}_k^{\text U},{\mathbf{I}_{\text{BD}}},{\mathbf{I}_{\text{IR}}}\} $, which is incorporated into the computation of $ {\cal F}$.  Moreover, instead of being pre-trained on a dataset, the learnable parameters are adaptively optimized in a task-driven manner using gradient descent on problem (P1), without requiring labeled data. As a result, each update of the learnable parameter set $\bold{\Theta} = \{ \Delta \bold{X},\bold{W}\} $ can be regarded as a form of implicit training embedded within the optimization process.

As can be seen from the Fig. \ref{Fig.3}, the constraints are achieved through different activation functions. For the maximum operator, i.e., $ \max ( \cdot , \cdot )$ in (\ref{eq15}), the rectified linear unit (ReLU) function can be adopted to replace them to facilitate the network training, and the Softmax activation function is employed to normalize the output of $\Delta {\bold{\widetilde X}^{(i)}}$. Moreover, we design a custom projection activation function that normalizes the output ${\bold{W}}$ according to the operation defined in (\ref{eq18}). Consequently,  (\ref{eq12}) can be implemented via a cumulative summation operation, which corresponds to the cumsum operator in PyTorch. Furthermore, a fixed learning rate that is too large causes oscillations, while too small slows convergence \cite{10659326}. In contrast, superior convergence speed and performance are demonstrated by Adam, primarily due to its adaptive learning rate mechanism, which mitigates sensitivity to the initial learning rate. The learnable parameters $\bold{\Theta}$ are updated by the Adam optimizer as follows

\begin{equation}\label{eq22}
{\bold{\Theta}^{(i)}} = {\bold{\Theta}^{(i - 1)}} + \eta  \cdot \text{Adam}({\nabla _\bold{\Theta}}{\cal F}),
\end{equation}
where $\eta $ denotes the initial learning rate. Upon completion of the training phase, the optimized variables $\bold X$ and ${\bold{W}}$ can be efficiently reconstructed from the learned network parameters and their associated transformations. 

The conventional neural network approach is typically trained offline with abundant samples to minimize the mean loss computed over the entire dataset. However, its performance may be limited by insufficient training data or convergence to local minima. In contrast, the proposed LGD method can be employed as a plug-and-play tool without requiring any pre-collected training dataset. Each update of the parameter set $\bold \Theta$ corresponds directly to a step in solving the original optimization problem, while the update direction and step size are still determined in accordance with the principles of GD \cite{9707491}. In addition, LGD exhibits strong interpretability, as its iterative variable update process is inherently explainable and can be seamlessly integrated with expert knowledge and prior information \cite{9805773}. Thanks to its lightweight network architecture, the proposed LGD method achieves low computational complexity, which can be characterized by $\mathcal{O}\left(I_G \big(KN^2M + K^2N\big)\right)$, where $I_G$ denotes the number of training iterations \cite{9707491}.

\begin{remark}
The solution to problem (P1) is achieved through iterative updates of the optimization variables, which are sensitive to the choice of initial values. To accelerate convergence and enhance performance, the initial value of $\Delta \bold{X}$ can be determined based on prior channel state information, while ${\bold{W}}$ must be initialized to satisfy the power constraint.
\end{remark}

\section{Proposed Two-stage Iterative Algorithm}
Although the proposed LGD adopts a lightweight network architecture with low computational complexity, solving problem (P1) via gradient descent can easily lead to convergence to local optima. This is mainly attributed to the periodic phase responses of the channel coefficients with respect to the antenna positions, as well as the strong coupling among the optimization variables, which leads to a large number of local minima\cite{wang2025model}. To further enhance performance, we propose a two-stage SCA-PSO algorithm. Given the strong coupling between the two optimization variables in (P1), the original problem is first decoupled into two subproblems, which are then solved in an alternating manner. Specifically, in the first stage, given the PA position matrix $\bold X$, the transmit beamforming matrix ${\bold{W}}$ is optimized by using the SCA method. In the second stage, the PSO method is developed to optimize $\bold X$ based on the optimized ${\bold{W}}$.

\subsection{Transmit Beamforming Optimization}
With fixed PA position matrix $\bold X$, the transmit beamforming subproblem can be written as

\begin{subequations}\label{eq23}
\begin{align*}
(\text{P2.1})\mathop {\max } \limits_{{\bold{W}}}\, &{\quad}\sum\limits_{k = 1}^K{\log _2}(1 + \frac{{{{\left| {\mathbf{h}_k^{eq}{\mathbf{w}_k}} \right|}^2}}}{{\sum {_{i = 1,i \ne k}^K{{\left| {\mathbf{h}_k^{eq}{\mathbf{w}_i}} \right|}^2} + \delta _k^2} }}) \\
&+ {\log _2}(1 + \frac{{{{\left| {(\mathbf{h}_k^{eq} + {\mathbf{f}_{b,k}}){\mathbf{w}_k}} \right|}^2}}}{{\sum {_{i = 1,i \ne k}^K{{\left| {(\mathbf{h}_k^{eq} + {\mathbf{f}_{b,k}}){\mathbf{w}_i}} \right|}^2} + \delta _k^2} }}).  \tag{23}\label{23}\\
{\rm{s.t.}} \, &(\text{\ref{11b}}), (\text{\ref{11c}})
\end{align*}
\end{subequations}
However, it can be seen that the objective function (\ref{eq23}) and constraint (\ref{11c}) are non-convex due to the quadratic terms, rendering the problem intractable to solve directly. To tackle this, we use the SCA algorithm to solve the problem (P2.1). By introducing the auxiliary variables 
${u_{1,k}} = {\mathbf{h}_{k}^{eq}}{\mathbf{w}_k}$, ${v_{1,k}} = \sum {_{i = 1,i \ne k}^K{{\left| {{ \mathbf{h}_{k}^{eq}}{\mathbf{w}_i}} \right|}^2} + \delta _k^2} $, $ {u_{2,k}} ={\left| {\left( {\mathbf{h}_k^{eq} + {\mathbf{f}_{b,k}}} \right){\mathbf{w}_k}} \right|^2}$, and $ {v_{2,k}} = \sum {_{i = 1,i \ne k}^K{{\left| {\left( {\mathbf{h}_k^{eq} + {\mathbf{f}_{b,k}}} \right){\mathbf{w}_i}} \right|}^2} + \delta _k^2}$, the objective function can be equivalently rewritten as ${\log _2}(1 + \frac{{{{\left| {{u_{1,k}}} \right|}^2}}}{{{v_{1,k}}}}) + {\log _2}(1 + \frac{{{{\left| {{u_{2,k}}} \right|}^2}}}{{{v_{2,k}}}})$. Note that the function $f(u,v) = {\left| u \right|^2}/v$ is convex with respect to $(u,v)$ for $v > 0$. The first-order Taylor expansion is used to construct the lower bound of the fractional quadratic term ${\left| {{u_{1,k}}} \right|^2}/{v_{1,k}}$ as follows:
\begin{equation}\label{eq24}
{t_{1,k}} \buildrel \Delta \over = 2{\mathop{\rm Re}\nolimits} \{ \frac{{\overline u _{1,k}^ * }}{{{{\overline v }_{1,k}}}}{u_{1,k}}\}  - \frac{{{{\left| {{{\overline u }_{1,k}}} \right|}^2}}}{{\overline v _{1,k}^2}}{v_{1,k}} \le \frac{{{{\left| {u_{1,k}^{}} \right|}^2}}}{{{v_{1,k}}}},
\end{equation}
where ${\overline u _{1,k}}$ and ${\overline v _{1,k}}$ denote the values of ${u_{1,k}}$ and ${u_{1,k}}$ at the previous iteration, respectively, and ${\rm{Re}}\left\{  \cdot  \right\}$ denotes the real part of the corresponding variable. Similarly, ${\left| {u_{2,k}^{}} \right|^2}/{v_{2,k}}$ is approximated by ${t_{2,k}} \buildrel \Delta \over = 2{\mathop{\rm Re}\nolimits} \{ \frac{{\overline u _{2,k}^ * }}{{{{\overline v }_{2,k}}}}{u_{2,k}}\}  - \frac{{{{\left| {{{\overline u }_{2,k}}} \right|}^2}}}{{\overline v _{2,k}^2}}{v_{2,k}}$. Moreover, based on (\ref{eq10}), we can introduce $D({P_0}|{P_1}) = L(\ln \rho  + {1 \mathord{\left/
 {\vphantom {1 \rho }} \right.
 \kern-\nulldelimiterspace} \rho } - 1)$, where $\rho  = {{({\gamma _b} + \delta _{{\rm{\text{IR}}}}^2)} \mathord{\left/
 {\vphantom {{({\gamma _b} + \delta _{{\rm{\text{IR}}}}^2)} {\delta _{{\rm{\text{IR}}}}^2}}} \right.
 \kern-\nulldelimiterspace} {\delta _{{\rm{\text{IR}}}}^2}}$. Accordingly, constraint (\ref{11c}) can be simplified as 

\begin{equation}\label{eq225}
\ln \rho  + {1 \mathord{\left/
 {\vphantom {1 \rho }} \right.
 \kern-\nulldelimiterspace} \rho } - 1 \ge {{2{\varepsilon ^2}} \mathord{\left/
 {\vphantom {{2{\varepsilon ^2}} L}} \right.
 \kern-\nulldelimiterspace} L}, 
\end{equation} which can be rewritten as $\frac{{{{\left\| {{f_{\text{BD},{\text{IR}}}}\mathbf{h}_{\text{BD}}^{eq}{\bold{W}}} \right\|}^2} + \delta _{{\rm{\text{IR}}}}^2}}{{\delta _{{\rm{\text{IR}}}}^2}} \le {\overline a _0}$ or $\frac{{{{\left\| {{f_{\text{BD},{\text{IR}}}}\mathbf{h}_{\text{BD}}^{eq}{\bold{W}}} \right\|}^2} + \delta _{{\rm{\text{IR}}}}^2}}{{\delta _{{\rm{\text{IR}}}}^2}} \ge {\overline a _1}$, where ${\overline a _0}$ and ${\overline a _1}$ are the two roots of the function $\ln \rho  + {1 \mathord{\left/
 {\vphantom {1 \rho }} \right.
 \kern-\nulldelimiterspace} \rho } - 1 = {{2{\varepsilon ^2}} \mathord{\left/
 {\vphantom {{2{\varepsilon ^2}} L}} \right.
 \kern-\nulldelimiterspace} L}$. It can be readily observed that ${\overline a _0} \le 1 \le \rho $, and therefore, constraint (\ref{11c}) can be reformulated as 

\begin{equation}\label{eq25}
{\left \| {{f_{\text{BD},{\text{IR}}}}\mathbf{h}_{\text{BD}}^{eq}{\bold{W}}} \right \|^2} \ge \left( {{{\overline a }_1} - 1} \right)\delta _{{\rm{\text{IR}}}}^2.
\end{equation}
To handle this non-convex constraint, the quadratic term $ {\left\| {{f_{\text{BD},{\text{IR}}}}\mathbf{h}_{\text{BD}}^{eq}{\bold{W}}} \right\|^2}$ is also approximated as $\varpi  \buildrel \Delta \over = 2{\rm{Re}}\left\{ \operatorname{Tr}({{{({f_{\text{BD},{\text{IR}}}}\mathbf{h}_{\text{BD}}^{eq}\bar {\bold{W}})}^H}{f_{\text{BD},{\text{IR}}}}\mathbf{h}_{\text{BD}}^{eq}{\bold{W}})} \right\} - {\left\| {{f_{\text{BD},{\text{IR}}}}\mathbf{h}_{\text{BD}}^{eq}\bar {\bold{W}}} \right\|^2}$ by using the first-order Taylor expansion, where $ {\rm{Re}}\left\{  \cdot  \right\}$ denotes the real part of the corresponding variable. Subsequently, the problem (P2.1) can be efficiently solved by utilizing the SCA algorithm. The approximated convex problem can be written as

\begin{subequations}\label{eq26}
\begin{align*}
(\text{P2.2})\mathop {\max } \limits_{X}\, &{\quad}\sum\limits_{k = 1}^K{\log _2}(1 + {t_{1,k}}) + {\log _2}(1 + {t_{2,k}})  \tag{26}\label{26}\\
{\rm{s.t.}} \,&{\left\| {{f_{\text{BD},{\text{IR}}}}\mathbf{h}_{\text{BD}}^{eq}{\bold{W}}} \right\|^2} \ge \left( {{{\overline a }_1} - 1} \right)\delta _{{\rm{\text{IR}}}}^2, \tag{26b}\label{26b}\\
&(\text{\ref{11b}}),
\end{align*}
\end{subequations}
which can be directly solved by the existing convex optimization tools such as CVX \cite{8630664}. The detailed SCA algorithm for solving the problem (P2.1) is summarized in \text{Algorithm 1}, where the initial ${\bold{W}}$ is randomly initialized in the feasible region.

\begin{algorithm}\label{alg1}
	\renewcommand{\algorithmicrequire}{\textbf{Input:}}
	\renewcommand{\algorithmicensure}{\textbf{Output:}}
	\caption{ SCA Algorithm for Solving (P2.1)}
	\label{alg1}
	\begin{algorithmic}[1]
		\STATE Initialize variables\  ${\bold{X}},  {\bar {\bold{W}}}$. Set iteration number $i=1$, the convergence accuracy   ${\varepsilon _{\text{1}}}$.
		\REPEAT
		\STATE Update $ {\bar u}_{1,k}$, ${{\bar v}_{1,k}}$, $ {\bar u}_{2,k}$, ${{\bar v}_{2,k}}$
		\STATE Update $ {\dot {\bold{W}}}$ = $ {\bold{W}} $ by solving problem (P2.2)
		\STATE Denote the objective value at $i$-th iteration as $ {\nu }^i$
		\STATE Set $i = i +1$
		\UNTIL $\left| {{{\nu }^{i + 1}} - {{\nu }^i}} \right| \le {\varepsilon _{{\rm{1}}}}$.
		
	\end{algorithmic}  
\end{algorithm}

\subsection{Pinching Beamforming Optimization}
With fixed transmit beamforming matrix ${\bold{W}}$, the pinching beamforming subproblem with respect to $\bold X$ can be formulated as

\begin{subequations}\label{eq27}
\begin{align*}
(\text{P2.3})\mathop {\max } \limits_{\bold{X}}\, &\sum\limits_{k = 1}^K {{R_k}},  \tag{27}\label{27}\\
{\rm{s.t.}} \,&(\text{\ref{11d}}), (\text{\ref{11e}}), (\text{\ref{26b}})
\end{align*}
\end{subequations}
which is challenging due to the ill-conditioned constraints and coupled variables in the objective function.   To address this issue, we adopt the PSO method to search for the optimal positions of the PAs, where PSO is a population-based stochastic optimization algorithm that mimics the social behavior of swarms to search for the optimal solution\cite{5601760}. While the exhaustive search guarantees the global optimum, its computational complexity is prohibitively high. PSO efficiently explores the solution space through swarm intelligence, achieving near-optimal performance with significantly reduced computational cost. Specifically, each waveguide is associated with a swarm, where each swarm consists of $Q$ particles. Taking a single waveguide as an example, we begin by randomly initializing $Q$ particles with positions $\mathbf{x}_q^{(t)} = {[x_{q,1}^{(t)},x_{q,2}^{(t)},\ldots,x_{q,M}^{(t)}]^T}$ and velocities $\mathbf{v}_q^{(t)} = {[v_{q,1}^{(t)},v_{q,2}^{(t)},\ldots,v_{q,M}^{(t)}]^T}$, $q = 1,\ldots,Q$ within the feasible search space. Here, $x_{q,m}^{(t)}$ and $v_{q,m}^{(t)}$ denote the position and the update velocity of $m$-th PA in the $q$-th particle during the t-th iteration, respectively. 

To satisfy constraint (\ref{11e}), all values of $x_{q,m}^{(t)}$ are restricted within the range $[0,{S_x}]$. Based on the PSO algorithm framework \cite{5601760,10818453}, each particle updates its position according to the current personal best position ${\widetilde x_{q,{\rm{pb}}}}$ and the swarm global best position ${\widetilde x_{g{\rm{b}}}}$. Accordingly, in $(t+1)$-th iteration, the update process for each particle’s velocity and position is formulated as follows

\begin{equation}\label{eq28}
\mathbf{v}_q^{(t + 1)} = {\omega _0}\mathbf{v}_q^{(t)} + {\omega _1}{c_1}({\widetilde x_{q,{\rm{pb}}}} - \mathbf{x}_q^{(t)}) + {\omega _2}{c_2}({\widetilde x_{{\rm{gb}}}} - \mathbf{x}_q^{(t)}),
\end{equation}
\begin{equation}\label{eq29}
\mathbf{x}_q^{(t + 1)} = \mathbf{x}_q^{(t)} + \mathbf{v}_q^{(t + 1)},
\end{equation}
where ${\omega _0}$ is the inertia weight that regulates the momentum of the particle and is defined as ${\omega _0} = {\omega _{\max }} - ({\omega _{\max }} - {\omega _{\min }}){t \mathord{\left/
 {\vphantom {t T}} \right.
 \kern-\nulldelimiterspace} T}$, where ${\omega _{\max }}$ and ${\omega _{\min }}$ represent the upper bound and lower bound of ${\omega _0}$, and $T$ denotes the maximum iteration number.
${\omega _1}$ and ${\omega _2}$ are random variables that follow a uniform distribution within the range $[0,1]$, introduced to improve the randomness of the search process to avoid premature convergence.  The parameters $c_1$ and $c_2$ act as the personal and global learning factors, respectively, regulating the weighting of the personal and global best positions to the velocity update.

\begin{algorithm}[t]
\caption{PSO Algorithm for Solving Problem (P2.3)}
\label{alg:pso}
\begin{algorithmic}[1]
\REQUIRE Initialized ${\bold{W}}$, $Q$, $N$, $M$,  $[0, S_x]$, $\mu$, $T$, $c_1$, $c_2$, $\omega_{\max}$, $\omega_{\min}$
\FOR{each waveguide $n = 1, 2,\ldots, N$}
    \STATE Randomly initialize the position $\mathbf{x}_q^{(0)} = [x_{q,1}^{(0)},\ldots, x_{q,M}^{(0)}]^T$ and velocity $\mathbf{v}_q^{(0)}$ for each particle $q = 1,\ldots, Q$
    \STATE Evaluate fitness $\mathcal{L}(\bold{X}_q^{(0)})$ for each particle using \eqref{eq30}
    \STATE Set personal best $\mathbf{\widetilde{x}}_{q,\mathrm{pb}} = \mathbf{x}_q^{(0)}$ and global best $\mathbf{\widetilde{x}}_{\mathrm{gb}} = \arg\max\limits_q \mathcal{L}(\bold{X}_q^{(0)})$
\ENDFOR
\FOR{iteration $t = 0$ to $T-1$}
    \STATE Update inertia weight: $\omega_0 = \omega_{\max} - (\omega_{\max} - \omega_{\min}) \cdot t/T$
    \FOR{each particle $q = 1,\ldots, Q$}
        \STATE Generate random numbers $\omega_1, \omega_2 \sim \mathcal{U}[0,1]$
        \STATE Update velocity $\bold{V}_q^{(t + 1)}$ by (\ref{eq28})
        \STATE Update position $\bold{X}_q^{(t+1)}$ by (\ref{eq29})
        \STATE Project $\bold{X}_q^{(t+1)}$ into feasible range $[0, S_x]$
        \STATE Evaluate fitness $\mathcal{L}(\bold{X}_q^{(t+1)})$ using \eqref{eq30}
        \IF{$\mathcal{L}(\bold{X}_q^{(t+1)}) > \mathcal{L}(\bold{\widetilde{X}}_{q,\mathrm{pb}})$}
            \STATE Update personal best: $\bold{\widetilde{X}}_{q,\mathrm{pb}} \leftarrow \mathbf{x}_q^{(t+1)}$
        \ENDIF
        \IF{$\mathcal{L}(\bold{X}_q^{(t+1)}) > \mathcal{L}(\bold{\widetilde{X}}_{\mathrm{gb}})$}
            \STATE Update global best: $\bold{\widetilde{X}}_{\mathrm{gb}} \leftarrow \bold{X}_q^{(t+1)}$
        \ENDIF
    \ENDFOR
\ENDFOR
Set the optimized antenna positions $\bold{X} = \bold{\widetilde{X}}_{\mathrm{gb}}$
\RETURN  $\bold X$
\end{algorithmic}
\end{algorithm}

In each iteration, the fitness value of the $q$-th particle is evaluated using equation (\ref{eq27}), based on its current position $\bold{X}_q$. To enforce constraints (\ref{11d}) and (\ref{26b}), an adaptive penalty method integrates the constraint violations into the objective function as penalty terms. The resulting penalized objective function is given by

\begin{equation}\label{eq30}
{\cal L}({\bold{X}_q}) = \sum\limits_{k = 1}^K {{R_k}({\bold{X}_q})}  - \mu \left| {{\cal P}({\bold{X}_q})} \right|,
\end{equation}
where ${\cal P}({\bold{X}_q})$ denotes the set of penalty terms associated with violations of the minimum detection error rate constraint (\ref{26b}) and the minimum PA spacing constraint (\ref{11d}). Specifically, $\mathcal{P}(\bold{X}_q)$ is defined as

\begin{align}\label{eq31}
\nonumber
{\cal P}({\bold{X}_q}) = \{ {x_{q,n,m}}&\left| {\{ {x_{q,n,m + 1}} - } \right.{x_{q,n,m}} < {d_{\min }},\forall n,m\} ,\\
&+D({P_0}|{P_1}) < 2{\varepsilon ^2}\} ,
\end{align}
where $\mu  > 0$ is a sufficiently large penalty factor that drives particles toward feasible regions. If a particle violates either constraint, the resulting fitness value $\mathcal{L}(\bold{X}_q)$ is penalized accordingly, potentially reducing it below zero to discourage infeasible solutions. As each particle is evaluated, its personal best and the global best positions are progressively updated until convergence is achieved. The detailed PSO algorithm for solving problem (P2.3) is summarized in Algorithm 2.

Based on the above analysis, the problem (P1) can be effectively solved using the proposed SCA-PSO algorithm, where the transmit beamforming matrix and the pinching position matrix are optimized in an alternating manner until convergence or the iteration limit is reached.

\subsection{Convergence and Complexity Analyses}
Since the proposed SCA-PSO algorithm operates in two stages, its convergence behavior depends on the performance of the SCA-based algorithm in the first stage and the PSO-based algorithm in the second stage. Denote by $f({\bold{W}^{(j)}},{\bold{X}^{(j)}})$ the objective function value at the $j$-th iteration. In this case, the following inequality holds:
\begin{equation}\label{eq32}
f({\bold{W}^{(j - 1)}},{\bold{X}^{(j - 1)}})\mathop  \le \limits^{(a)} f({\bold{W}^{(j)}},{\bold{X}^{(j - 1)}})\mathop  \le \limits^{(b)} f({\bold{W}^{(j)}},{\bold{X}^{(j)}}),
\end{equation}
where the inequality in (a) arises from the fact that problem (P2.2) is optimally resolved per iteration, and its optimal objective value constitutes a lower bound for that of problem (P2.1). The inequality (b) holds because the fitness value of the global best position is non-decreasing throughout the iterations of Algorithm 2, i.e., $ {\cal L}({\bold{X}}_q^{(t+1)}) \ge {\cal L}({\bold{X}}_q^{(t)}) $. If the updated pinching-antenna position obtained from the PSO algorithm fails to yield an improvement in sum rate, the previous position is preserved \cite{11131179}. Moreover, since the objective value of problem (P2.3) is bounded above, the sequence of best fitness values converges, ensuring the convergence of the PSO-based algorithm.

The computational complexity of the proposed SCA-PSO algorithm is primarily determined by the subproblems (P2.2) and (P2.3). Specifically, as indicated in \cite{9133435,9234527}, the subproblem (P2.2) takes the form of a second-order cone programming (SOCP) problem, which can be effectively resolved via an interior-point method, incurring a computational complexity of $ {\log (\frac{1}{{{\varepsilon _1}}}){{({K^2}N)}^{3.5}}}$. To address the subproblem (P2.3), the PSO algorithm used for updating $\bold X$ has a computational complexity of ${\mathcal O}\left( {TQMN} \right)$. Therefore, the overall complexity of the proposed algorithm is expressed as $ {\mathcal O}\left( {TQMN + \log (\frac{1}{{{\varepsilon _1}}}){{({K^2}N)}^{3.5}}} \right)$.

\begin{figure*}[htbp]
    \centering
    \subfloat[\footnotesize \text{}]{%
        \includegraphics[width=0.49\textwidth]{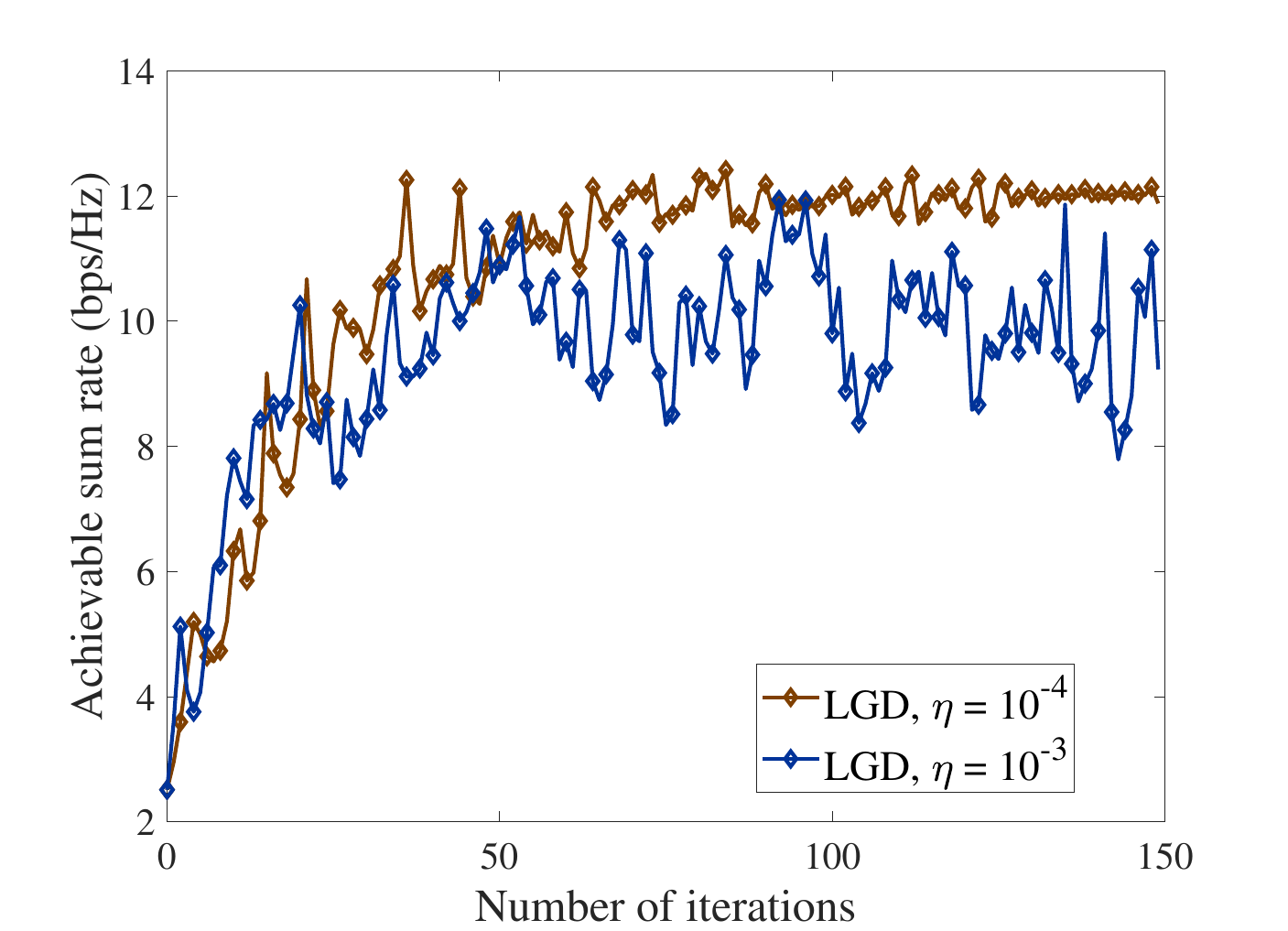}
    }
    \subfloat[\footnotesize \text{}]{%
        \includegraphics[width=0.49\textwidth]{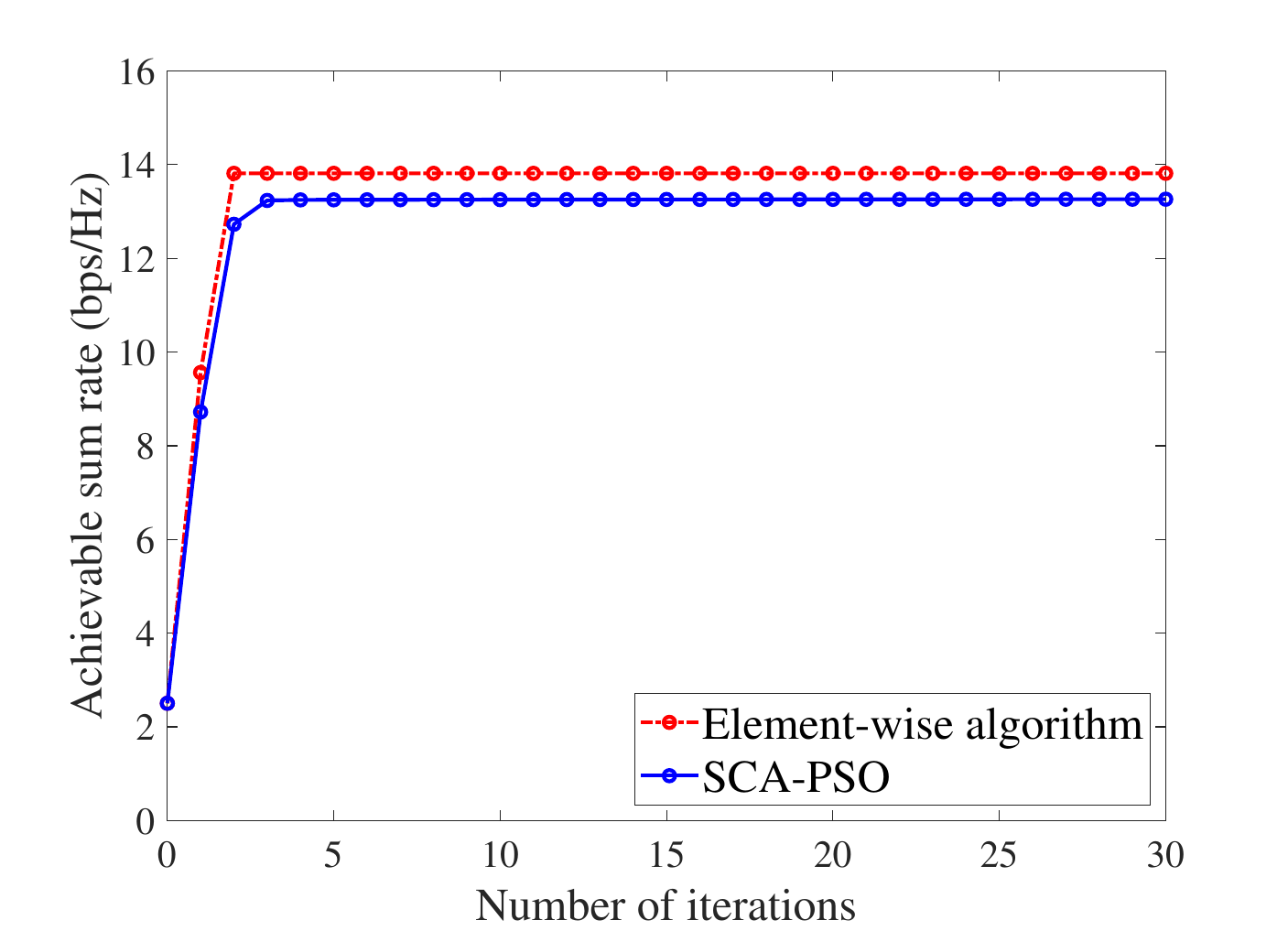}
    }
    \caption{Convergence behaviour of the proposed algorithms with $K=N=2$, $M=3$, and $P_{\text{max}}=30\,\text{dBm}$. (a) shows the convergence behavior of the proposed LGD algorithm under the learning rate $\eta  = \{ {10^{ - 4}},{10^{ - 3}}\}$. (b) shows comparisons of the convergence behavior of the proposed SCA-PSO with the element-wise algorithm.}
    \label{fig_4}
\end{figure*}

\section{Numerical Results}
This section specifies the simulation setup and provides numerical results to demonstrate the effectiveness of the proposed PASS-enabled SR design and associated algorithms. For consistency with prior work, the system parameters are adopted from \cite{xu2025join} and \cite{10896748}. Specifically, the operating frequency is set to $f=28 \,\text{GHz}$, the ma transmit power, the noise power is  $\delta _k^2 = \delta _{\text{IR}}^2 =  - 80 \, \text{dBm}$, and the effective index of the waveguide is $n_{\text{eff}}=1.4$. We assume that the numbers of PRs and chains/waveguides are $ N = K = \{ 2,4\} $. Each waveguide is equipped with $M=3$ PAs, and the heights of all PAs are fixed at $z^{\text{PA}}=5 m$.  Both the PRs and the IR are randomly deployed within a rectangular area of size ${S_x} \times {S_y} = 30 \times 4~\text{m}^2$, while the BD is placed at a fixed location of $(5, 2, 2) \, \text{m}$. The waveguides are uniformly distributed along the y-axis (vertical direction) with a consistent interval of ${{{S_{\rm{y}}}} \mathord{\left/
 {\vphantom {{{S_{\rm{y}}}} N}} \right.
 \kern-\nulldelimiterspace} N}$ meters and the minimum spacing of PA positions is set to $d_{\min}$ = 0.1 m. The symbol duration ratio of the primary to the secondary transmission is set as$L = 60$, and the detection probability threshold is $\varepsilon = 0.95$. The convergence tolerance and penalty factor in Algorithm 1 are set to $\varepsilon_1 = 10^{-3}$ and $\mu=10$, respectively. The swarm size and the maximum number of iterations in Algorithm 2 are set to $Q=30$ and $T=100$, respectively. For the LGD method, the penalty parameter is set to $\xi  = {10^3}$, and the learning rate of the Adam optimizer is $\eta=10^{-4}$, while the remaining Adam hyperparameters follow the default settings. The proposed schemes are implemented using the PyTorch library and MATLAB R2024a on a platform equipped with an Intel Core i5-14600K processor and 64 GB RAM. Numerical results are computed as the average over 100 independently generated channel realizations.

For performance comparison, the following benchmark schemes are considered: 1) \textbf{Element-wise algorithm}: Each antenna position $x_{n,m}$ is optimized individually using a one-dimensional search to obtain a near-optimal solution\cite{wang2025model}.  Specifically, denote by ${{\cal D}_x}$ a uniform sampled feasible grid over the interval $[0,{S_x}]$. The optimization with respect to $ x_{n,m}$ can be formulated as $\mathop {\max }\limits_{{x_{n,m}} \in {{\cal D}_x}} {\rm{ }}\sum {_{k = 1}^K} {R_k}$, where the suboptimal solution $x_{n,m}^ * $ is obtained by searching over the feasible discrete grid ${{\cal D}_x}$, subject to the constraints (\ref{11c})-(\ref{11e}). 2) \textbf{Fixed-PA}: The PAs are uniformly distributed along the x-axis within each waveguide, with an inter-element spacing ${{{S_x}} \mathord{\left/ {\vphantom {{{S_x}} M}} \right. \kern-\nulldelimiterspace} M}$. 3) \textbf{Massive MIMO}: The massive MIMO BS with the hybrid beamforming architecture is positioned at the origin point, where $N$ RF chains are deployed, each connected to $M$ antennas through phase shifters. 4) \textbf{Conventional MIMO}: The MIMO BS equipped with $N$ antennas is placed at the origin point, each connected to a dedicated RF chain.

\begin{figure}[t]
\centering
\includegraphics[width=0.49\textwidth]{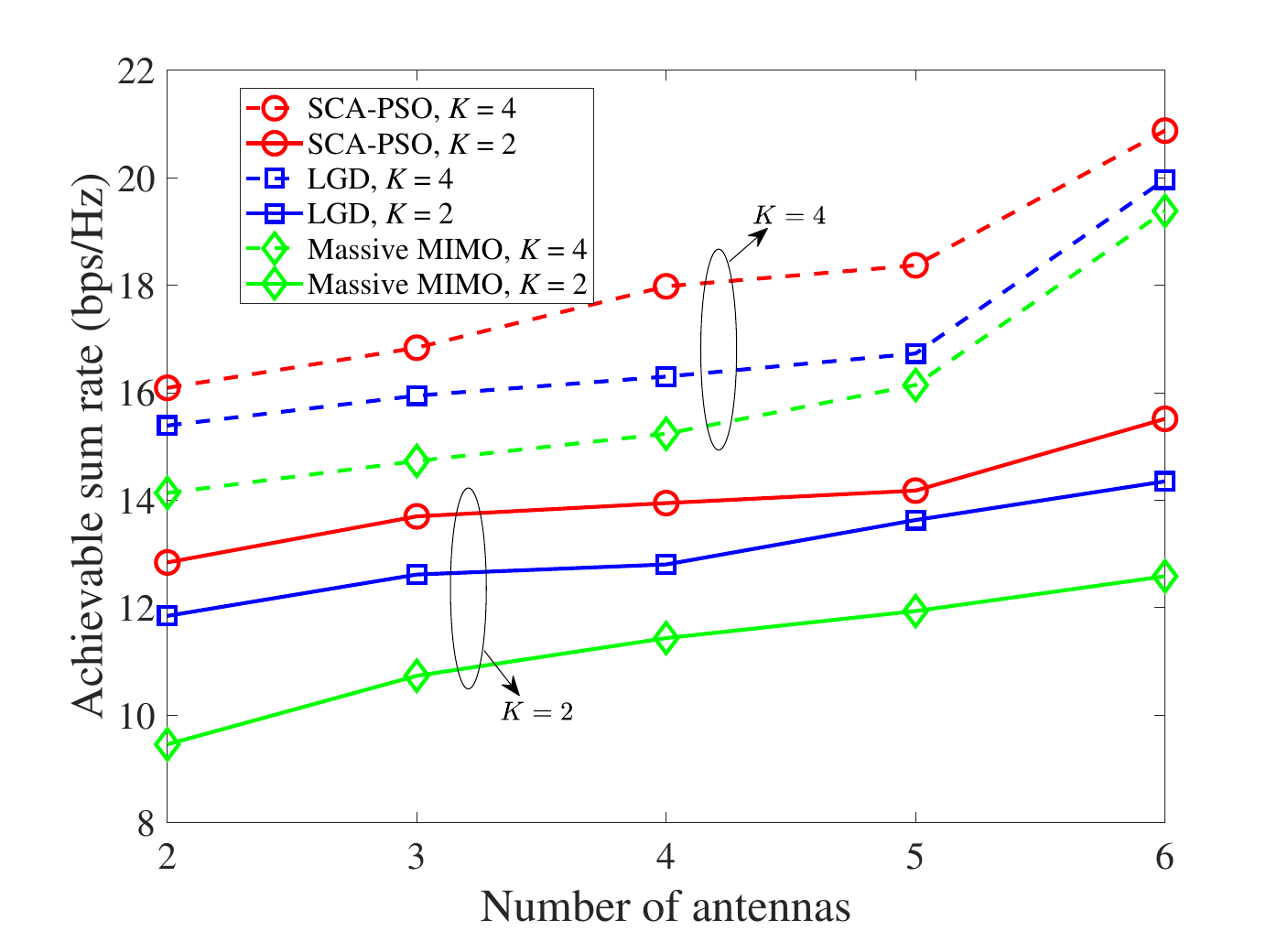}
\caption{ Achievable sum rate versus the number of antennas under different $K$.}
\label{fig_5}
\end{figure}

Fig. \ref{fig_4} presents the convergence characteristics of the proposed algorithms. As shown in Fig. \ref{fig_4}(a), the LGD algorithm demonstrates different behaviors under varying initial learning rates. The results show that although a larger initial learning rate facilitates faster initial ascent during the early iterations, it often suffers from pronounced fluctuations and is more difficult to converge to suboptimal solutions. In contrast, the LGD algorithm with an initial learning rate $\eta=10^{-4}$ leads to a more stable convergence trajectory and ultimately achieves a higher sum rate. Fig. \ref{fig_4}(b) compares the convergence performance of the proposed SCA-PSO algorithm against the element-wise benchmark. It can be observed that both algorithms converge rapidly within a few iterations, and the achievable sum rate gradually stabilizes as the number of iterations increases. The element-wise algorithm achieves a slightly higher achievable sum rate due to its exhaustive search over antenna positions. Specifically, the antenna position matrix $\bold{X}$ is updated using the element-wise algorithm, whose computational complexity is expressed as $O({I_{{\rm{iter}}}}UMNK)$ \cite{wang2025model}, where $I_{\rm{iter}}$ denotes the number of iterations. Moreover, Table \ref{tab1} compares the computational complexity and running time of different algorithms. It can be observed that the element-wise algorithm requires a significantly longer runtime than the proposed SCA-PSO algorithm. This is because it performs a dense position search over the entire feasible region, i.e., $ U \gg TQ$, resulting in prohibitively high computational complexity. The SCA-PSO algorithm provides a much more computationally efficient solution while maintaining competitive performance. The LGD method requires the least runtime due to its lightweight network architecture.

Fig. \ref{fig_5} compares the achievable sum rates of different approaches as the number of active pinches/antennas per waveguide (or RF chain) increases. For all considered schemes, the sum rate improves with the number of active antennas, confirming the benefits of increased spatial degrees of freedom for beamforming and interference mitigation. In both the $K=2$ and $K=4$ user cases, the sum rate also increases with $K$ due to spatial multiplexing gains. Leveraging PASS, the proposed SCA-PSO and LGD schemes achieve substantial sum-rate improvements over the baselines, primarily by reconfiguring large-scale path loss through flexible position optimization. When the maximum range of each waveguide is fixed, the performance gap between the LGD method and the massive MIMO scheme gradually narrows as the number of antennas increases.

\begin{figure}[t]
\centering
\includegraphics[width=0.49\textwidth]{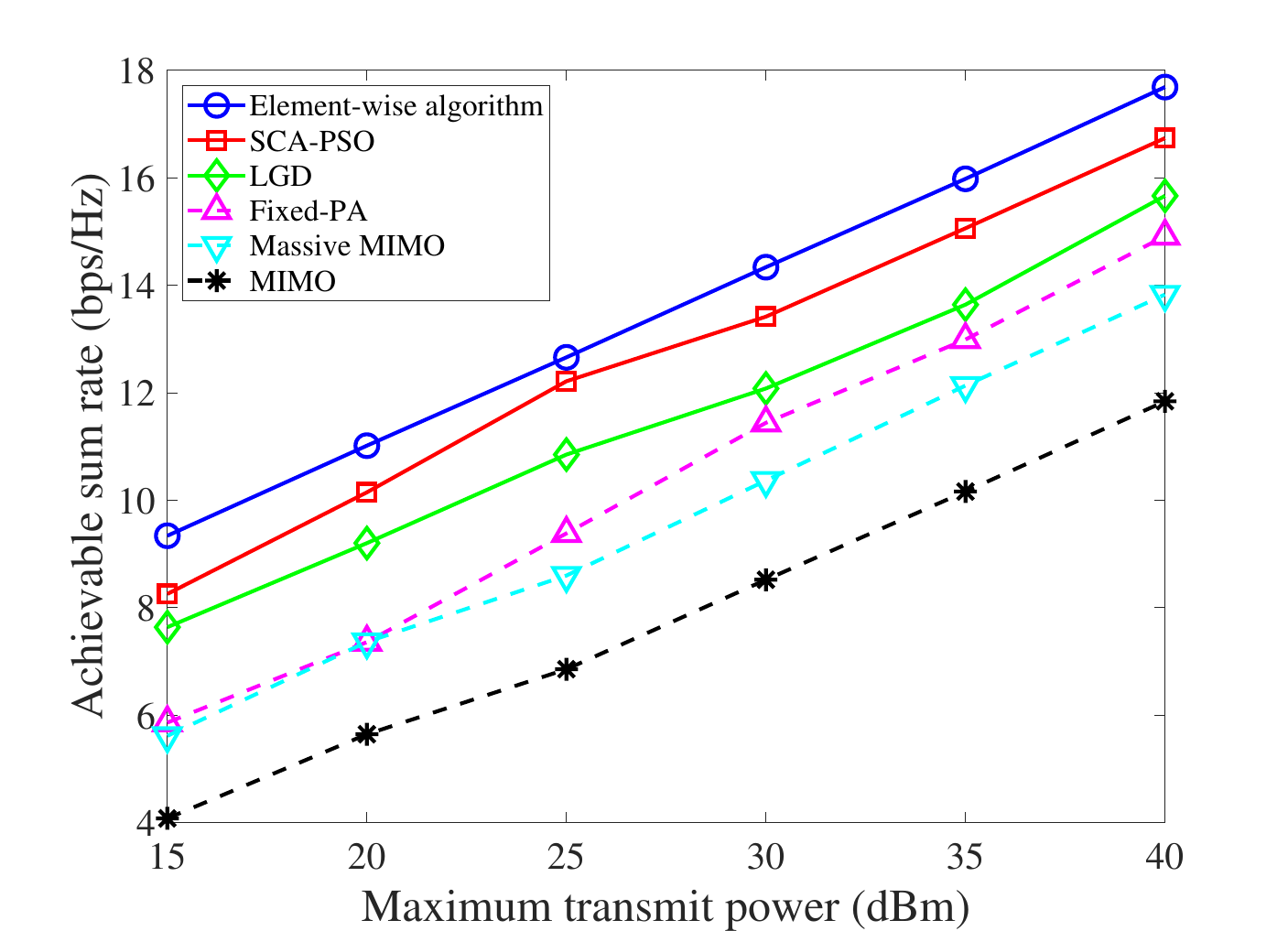}
\caption{ Achievable sum rate versus the maximum transmit power.}
\label{fig_6}
\end{figure}

\begin{table}[t]
\caption{Computational Complexity and Running Time Comparison}
\centering
\footnotesize
\label{tab1}
\begin{tabular}{|c|c|c|}
\hline
\makecell{Scheme} & \makecell{Computational\\ Complexity} & \makecell{Running\\ Time (s)} \\
\hline
LGD & $\mathcal{O}(I_G(KN^2M + K^2N))$ & 1.076\\
\hline
SCA-PSO & $\mathcal{O}(TQMNK + \log(1/\varepsilon)(K^2N)^{3.5})$ & 16.43\\
\hline
Element-wise & $\mathcal{O}(I_{\rm{iter}} UMNK + \log(1/\varepsilon)(K^2N)^{3.5})$ & 86.315\\
\hline
\end{tabular}
\end{table}

Fig. \ref{fig_6} presents the achievable sum rates of different algorithms under varying maximum transmit power $P_{\max}$ . For all considered schemes, the achievable rate increases with $P_{\max}$, as higher transmit power enhances the received signal strength and improves the SINR. The proposed SCA-PSO and LGD algorithms consistently outperform the conventional MIMO and massive MIMO baselines across the entire transmit power range. This confirms the effectiveness of the proposed optimization framework in alleviating large-scale path loss significantly. Among them, SCA-PSO achieves higher performance than LGD by leveraging its stochastic–deterministic search strategy to obtain a high-quality suboptimal solution. In contrast, LGD is more prone to becoming trapped in local optima near the initial point, which can lead to performance degradation. While the element-wise algorithm attains the highest sum rate overall, it comes with significantly higher computational complexity due to its exhaustive search over individual antenna elements, which may limit its practical deployment. In contrast, the proposed SCA-PSO strikes a favorable balance between performance and computational efficiency.

\begin{figure}[t]
\centering
\includegraphics[width=0.49\textwidth]{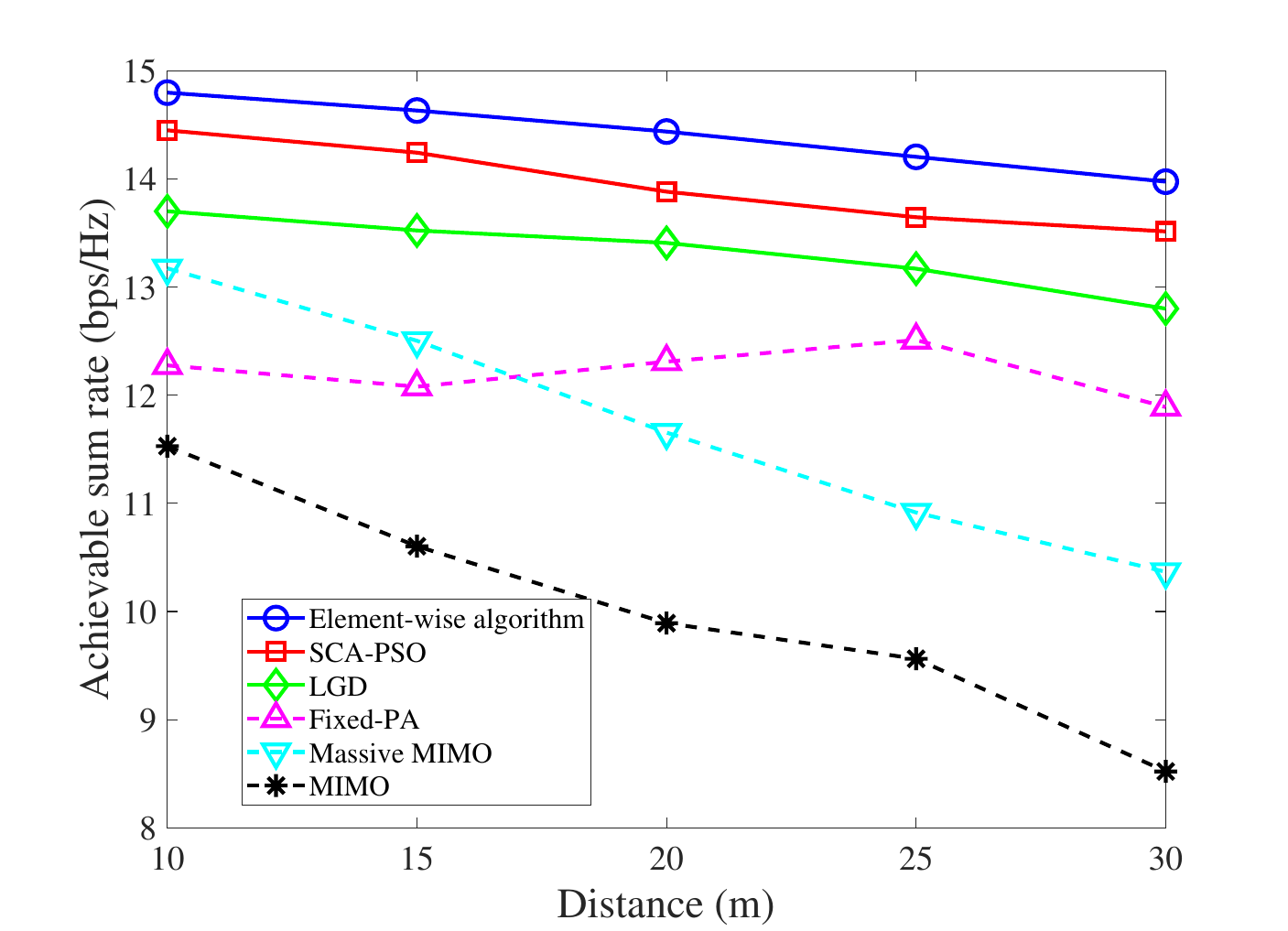}
\caption{ Achievable sum rate versus the range distance $S_x$.}
\label{fig_7}
\end{figure}

In Fig. \ref{fig_7}, we compare the achievable sum rate of different schemes as the distance $S_x$ increases from 10 m to 30 m. It can be observed that all schemes experience performance degradation with the increase of distance. However, the sum rate achieved by PASS exhibits only a marginal decrease in the sum rate. This degradation is caused by increasing channel attenuation on the BD-to-PRs/IR links, while the path loss on the dominant primary transmission paths from the BS is effectively mitigated through flexible placement of the PAs. In contrast, the performance of MIMO and Massive MIMO decreases rapidly as the distance increases from 10 m to 30 m, with a sum rate reduction of approximately 21\%, mainly due to the free-space path loss effect. Massive MIMO achieves competitive performance at shorter ranges due to its hybrid beamforming architecture, and the Fixed-PA scheme exhibits fluctuating performance as its fixed antenna positions cannot adapt to varying link conditions. Thanks to the flexibility of PA deployment, PASS not only minimizes the long-distance path loss on the primary transmission links but also alleviates the adverse effects of double fading in backscattering links.

\begin{figure}[t]
\centering
\includegraphics[width=0.49\textwidth]{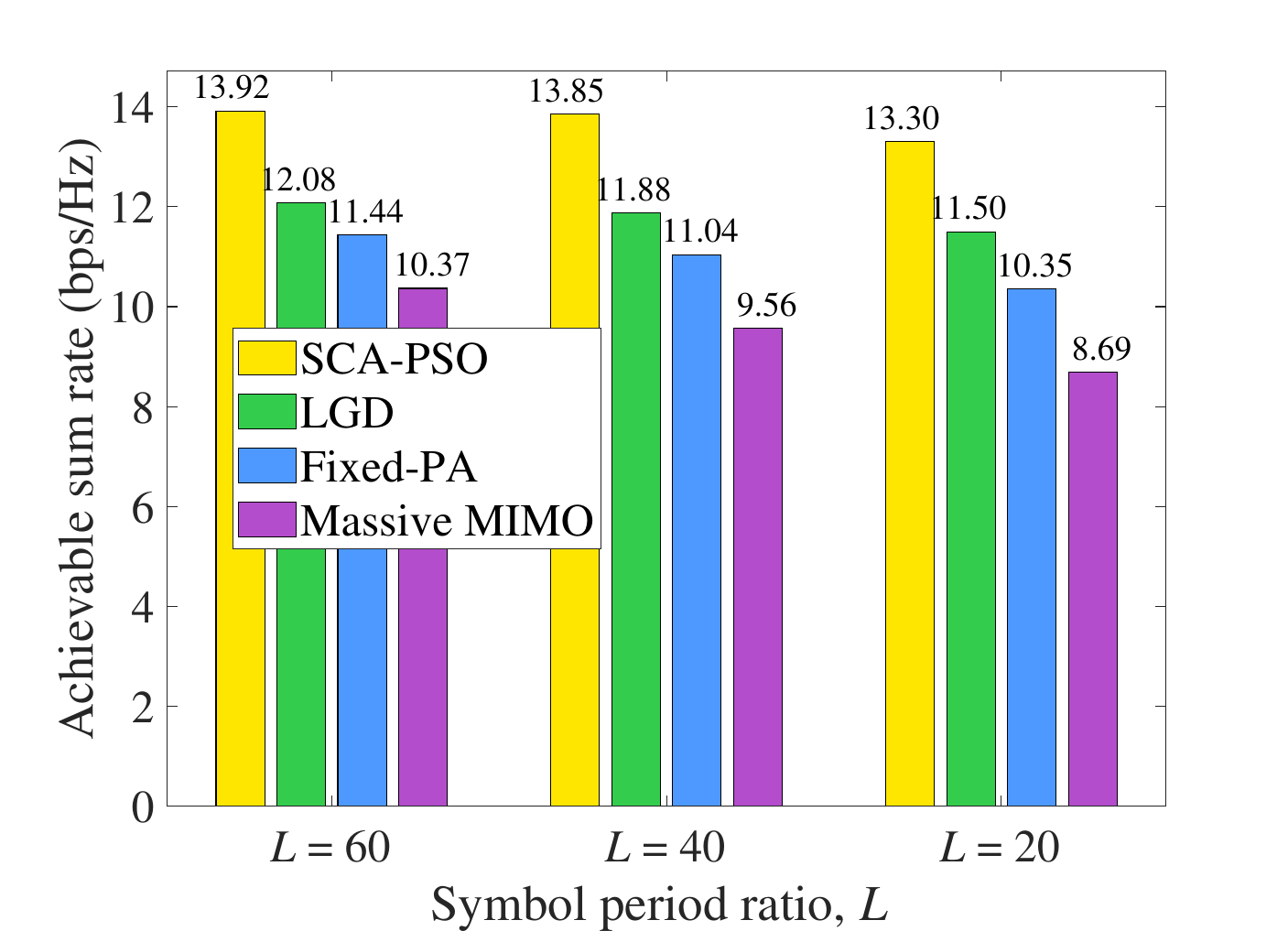}
\caption{ Achievable sum rate versus symbol period ratio, $\it{L}$.}
\label{fig_8}
\end{figure}

We investigate the impact of the symbol period ratio $L$, as illustrated in Fig. \ref{fig_8}. The results indicate that the performance of all schemes degrades as $L$ decreases. This is because a smaller $L$ implies fewer information symbols $s(l)$ within one $c$’s symbol period, which in turn deteriorates the decoding performance of the secondary signal and ultimately limits the overall system performance. As observed, the proposed scheme achieves a higher sum rate compared to the fixed-PA scheme, demonstrating the effectiveness of optimizing the positions of the PAs. Moreover, in this context, PASS demonstrates a stronger capability in mitigating this issue, resulting in less performance degradation compared to conventional antenna systems. This also indicates the potential of PASS as an enabling technology for PSR systems, which is worth further exploration in future work.

In Fig. 9, we evaluate the impact of transmit SNR on the achievable sum rate. As expected, the sum rate of all schemes increases with higher SNR, owing to the improved signal strength relative to noise. Specifically, for the PASS-based system with the proposed schemes, when the transmit SNR is 65 dBm, the SCA-PSO and LGD algorithms achieve a 65.6\% and 55.9\% gain in sum rate, respectively, compared to conventional massive MIMO. This performance improvement is attributed to the flexibility of PASSs in dynamically reconfiguring pinching positions along the waveguide, thereby enabling the formation of strong and reliable LoS links. Furthermore, the performance gap between PASS-based schemes and massive MIMO widens as SNR increases. This is because, at low SNR, noise interference not only degrades the achievable data rate but also increases the detection error probability of the secondary signal.

\begin{figure}[t]
\centering
\includegraphics[width=0.49\textwidth]{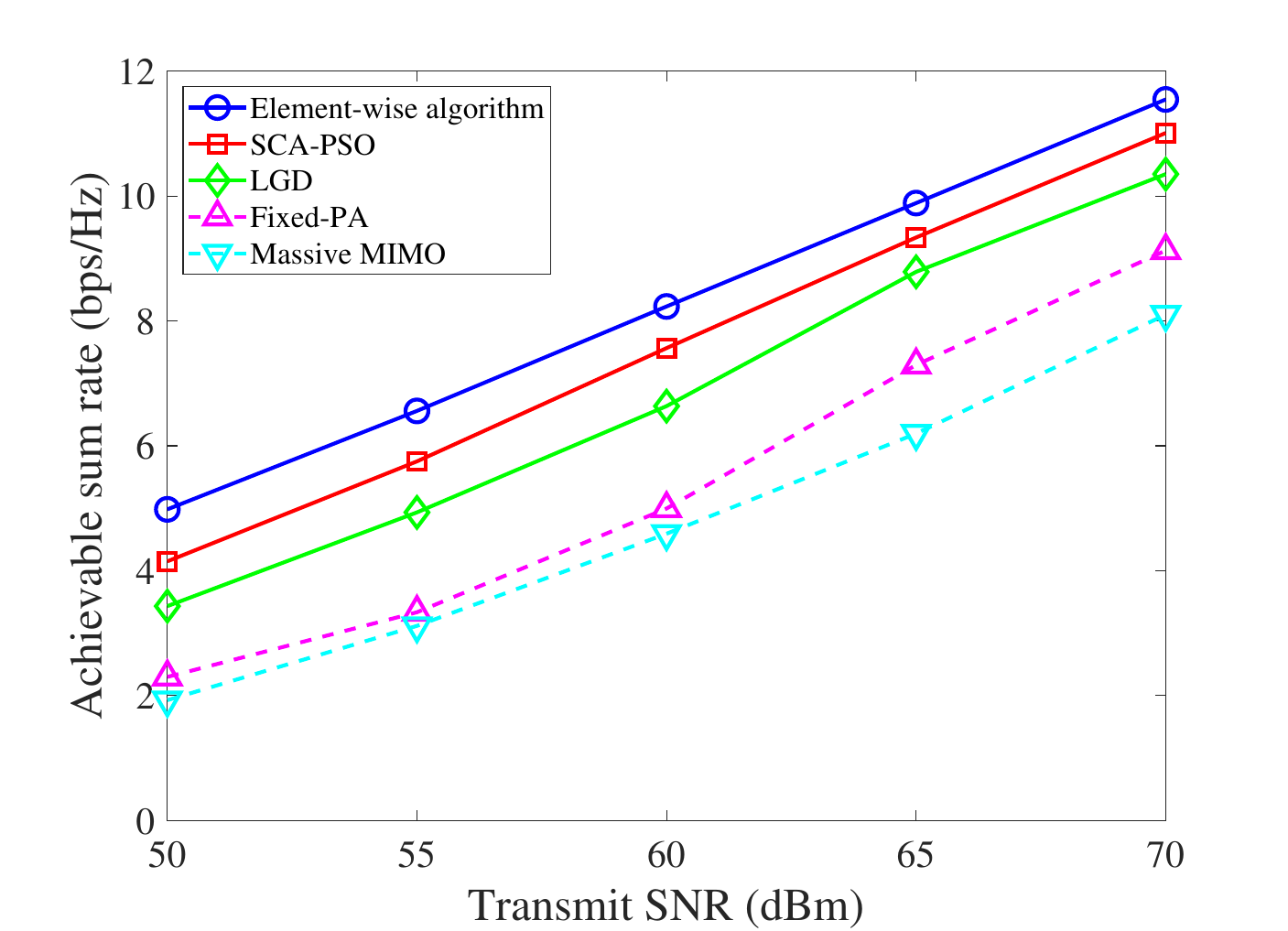}
\caption{ Achievable sum rate versus the transmit SNR.}
\label{fig_9}
\end{figure}

\section{Conclusion}
In this paper, we have proposed a PASS-enabled downlink SR framework that jointly optimizes transmit and pinching beamforming to enhance both primary and secondary transmissions, where a PASS-based BS equipped with multiple waveguides, acting as the PT, serves both the IR and PRs with the aid of the BD. To maximize the sum rate, we have formulated a joint optimization problem for transmit and pinching beamforming, subject to the detection error probability constraint at the IR and the feasible deployment region of the PAs. To address the highly coupled and nonconvex nature of this problem, we have developed two solution approaches. The first is a low-complexity LGD method, which leverages end-to-end learning to transform parameters and incorporate constraints, thereby efficiently solving the constrained problem based on the principle of gradient descent. To further enhance performance, we introduced the SCA-PSO-based approach, where the transmit beamforming was optimized via SCA, followed by pinching beamforming optimization through a PSO-based search over feasible PA positions.  Simulation results have verified that the proposed PASS framework achieves notable performance gains over conventional fixed-antenna and massive MIMO schemes under strict detection error constraints. In particular,  SCA-PSO attained performance close to the element-wise benchmark while significantly reducing computational complexity.

Future work may explore PA-enabled PSR scenarios, which fully leverage the path-loss reconfiguration capability of PAs to enhance the efficiency of backscatter communication. In addition, the design of multi-antenna receiver architectures and the adoption of higher-order modulation schemes represent promising research directions.

\bibliographystyle{IEEEtran}
\bibliography{ref}

\vfill

\end{document}